\documentclass[useAMS,usenatbib]{mn2e}
\usepackage{dcolumn}
\usepackage{bm}
\usepackage{amsmath,amsfonts,amssymb}
\usepackage{natbib}
\usepackage{color,graphicx}
\usepackage{epsfig}
\usepackage{subfigure}
\usepackage{enumitem}
\usepackage{units}
\usepackage{mathrsfs}
\usepackage{hyperref}
\usepackage{verbatim}

\begin{document}
\setlength{\pdfpageheight}{\paperheight}
\setlength{\pdfpagewidth}{\paperwidth}
\title{Ultrarelativistic Bondi--Hoyle Accretion I: Axisymmetry}
\author[A.J.~Penner]{A.J.~Penner\thanks{Andrew-Jason.Penner@obspm.fr} \\
$^1$Laboratoire Univers et Th\'eories, CNRS/Observatoire de Paris/Universit\'e Paris Didrot, 5 Place Jules Jannsen,92195 Meudon ,France\\ 
$^2$School of Mathematics, University of Southampton, Southampton SO17 1BJ, UK}
%
\date{May 10 2012; Revised XX; \LaTeX-ed \today}
\maketitle

\begin{abstract}
An ultrarelativistic relativistic study of axisymmetric Bondi--Hoyle accretion onto a moving Kerr black hole is presented. The equations of general relativistic hydrodynamics are solved using high resolution shock capturing methods. In this treatment we consider the ultrarelativistic limit wherein one may neglect the baryon rest mass density. This approximation is valid in the regime where the internal energy of the system dominates over the rest mass energy contribution from the baryons. The parameters of interest in this study are the adiabatic constant $\Gamma$, and the asymptotic speed of the fluid, $v_\infty$. We perform our simulations in three different regimes, subsonic, marginally supersonic, and supersonic, but the primary focus of this study is the parameter regime in which the flow is supersonic, that is when $v_\infty \ge c_{s}^{\infty}$. As expected from previous studies the supersonic regimes reveal interesting dynamics, but even more interesting is the presence of a bow shock in marginally supersonic systems. A range of parameter values were investigated to attempt to capture possible deviations from steady state solutions, none were found. To show the steady state behaviour of each of the flows studied we calculate the energy accretion rates on the Schwarzschild radius. Additionally, we also find that the accretion flows are dependent on the location of the computational boundary, that if the computational boundary is located too close to the black hole the calculated flow profiles are marred with numerical artifacts. This is a problem not found in previous relativistic models for ultrarelativistic hydrodynamic systems.
\end{abstract}
\begin{keywords}
accretion, accretion discs -- black hole physics -- hydrodynamics -- relativity -- shock waves -- methods: numerical
\end{keywords}
\section{Introduction}
During the radiation dominated era the universe is thought to be comprised of radiation fluid, that is to say a fluid with internal energy which is sufficiently large that the rest mass density is negligible \citep{Miller:2005,Musco:2005,Musco:2008}. During this phase primordial black holes (PBH) would be created \citep{Carr:1974} and grow due to the accretion of this radiation fluid \citep{Sivaram:2001}. Over the last several decades there has been interest in the formation of primordial black holes (PBH). In particular there has been focus on the question about the growth rate of the PBH. Recent studies of primordial black holes by \citet{Custodio:2002} suggest that their accretion rates are relatively low, so that the mass of the black holes does not increase much through the accretion process. 

Previous studies used a simplified Bondi-type model (stationary accretor) with an ideal gas equation of state to show that the accretion rates would not be sufficient to appreciably modify the mass of the PBH if the horizon of the PBH is smaller than the cosmological horizon \citep{Zeldovitch:1967,Carr:2010}. In the event that the PBH is on the order of the cosmological horizon the black hole would expand at the same rate as the universe \citep{Zeldovitch:1967,Carr:2010}. However, as \citet{Carr:2010} explain, if the latter were true we would expect the existence of black holes that are several orders of magnitude larger than the black holes found in active galactic nuclei. Further studies of PBH lead to proof that they cannot expand at the same rate as the universe as shown in \citet{Carr:1974}.

More recent research continues the debate over the existence of observable PBHs. Some authors show that the calculated accretion rates for simplified models are sufficiently high to transform a PBH to solar mass black holes \citep{Sivaram:2001}. As cited above, others make claims that when more complicated accretion models are considered, the mass accretion rates for both stationary and moving PBHs are insufficient to appreciably modify the mass of the PBH except in very high Lorentz factor models \citep{Custodio:2002,Horvath:2005}. We show, using our own simplified model, that the accretion rates of moving PBH is insufficient for the PBH to reach solar masses, in agreement with the conclusions of \citet{Custodio:2002}.

Traditional use of perfect radiation fluids is seen in cosmological studies, where one would expect the equation of state to be \citep{Miller:2005},
\begin{equation}
 P=\frac{1}{3}\left(\rho-\frac{\Lambda}{2\pi}\right),
\end{equation}
where $P$ is the pressure, $\rho$ is the mass-energy density, and $\Lambda$ is the cosmological constant.
If we generalize this relation between the fluid pressure, $P$, and energy density $\rho$, we have\footnote{We used geometric units so that the speed of light is unity.},
\begin{equation}
 P=\left(\Gamma-1\right)\left(\rho-\frac{\Lambda}{2\pi}\right).\label{eq:EoS}
\end{equation}
This equation of state gives us a speed of sound,
\begin{equation}
 c_s^2 = \frac{dP}{d\rho} = \Gamma-1.\label{eq:SoS}
\end{equation}
Due to relativistic limits on the speed of sound we restrict our attention to $\Gamma<2$. Furthermore, we are interested in treating the background matter as a fluid so we also restrict our attention to matter with $\Gamma>1$. When $\Gamma=1$ the background fluid becomes a pressure-less dust.

The equation of state \eqref{eq:EoS} is valid in the very early universe, $\sim10^{-5}$s after the Big Bang close to the time of the cosmological quark-hadron transition. This is deep inside the radiation-dominated era \citep{Miller:2005}. During this time we expect the temperatures to be on the order of $10^{12}$K, mass-energy densities and horizon scales on the order of neutron star central densities, $\sim10^{15} \rm{g}/\rm{cm}^{3}$, and radius $\sim10$km respectively. The energy density of the early universe was dominated by relativistic particles and radiation. Thus equation of state \eqref{eq:EoS} is valid for any part of the radiation dominated phase, up to $\sim10^{11}$s after the Big Bang \citep{Miller:2005}. Consequently, during this time, we expect our equation of state to be valid, even on scales close to the black hole size $\sim10^{15}$g.

Our model assumes that the background fluid has uniform density. The fluid itself is not sufficiently dense to experience an appreciable self-gravity. We assume that for the duration of these simulations that the background spacetime is stationary. This assumption is in agreement with results from \citet{Custodio:2002}, and allows us to treat the accretion problem using relativistic Bondi \citep{Michel} and relativistic Bondi--Hoyle \citep{Penner:2011,FI1,FI2,FI3,PSST,Petrich:1988} accretion models. Furthermore, this paper focuses on axisymmetric accretion onto an axisymmetric black hole, the next paper in this series studies a similar accretion system in an infinitely-thin disc model.

In previous studies by \citet{FI1}, an ultrarelativistic system was investigated and determined to reach a steady state. However, their physical model specified the speed of sound in the background fluid to be $c_s = \sqrt{\Gamma-1}$, while maintaining the usual ideal fluid enthalpy per unit rest mass, 
\begin{equation}
 h = 1 + \epsilon + P/\rho_0.
\end{equation}
Our model of the background fluid's equation of state enforces the ultrarelativistic model such that the enthalpy density has the form,
\begin{equation}
 h = \rho +P.\label{eq:enthalpy}
\end{equation}
In our picture $\rho$ does not include the rest mass density.

For simplicity, our model assumes that the cosmological constant, $\Lambda$, is zero. While this constant is important for the dynamic study of gravitational collapse, in the early universe, $\Lambda$ does not change the speed of sound \eqref{eq:SoS} and thus does not change the characteristic parameters in the system. Consequently, the cosmological constant is not expected to be dynamically important in our fixed spacetime model.

To be consistent with ``standard'' radiation fluids we primarily investigate fluids with adiabatic index $\Gamma=4/3$. However, we extend our model to investigate a range of adiabatic constants and found similar behaviour in all cases. The parameters investigated may be found in table \ref{table:1}. For an ultrarelativistic fluid with $\Gamma=2$ we refer the reader to \citet{Petrich:1988}, where a closed form solution is presented. We investigate three regions of parameter space, the subsonic regime $v_{\infty} < c_s$, the marginally supersonic regime $v_\infty \gtrsim c_s$, and the supersonic regime $v_\infty > c_s$. As expected from previous studies the supersonic regimes contain the most interesting dynamics.

Another interest of previous relativistic Bondi--Hoyle studies \citep{FI1,FI2,FI3,PF1,Donmez:2010} focused on the search for an instability that would develop as the fluid accretes onto the black hole \citep{FI1}. This instability is known as the ``flip-flop'' instability and is realized by the motion of the tail shock that rapidly moves from one side of the black hole to the other, eventually moving to the upstream side of the black hole destroying the stable accretion flow. \citet{FI1} show that the evolution of the axisymmetric relativistic Bondi--Hoyle setup result in steady-state accretion rates for the parameters investigated. Since any instabilities would be strongest on the event horizon, we calculate the accretion rates on the event horizon of the black hole. We further show, for the duration of our simulations, that instabilities were not found.

To parameterise our fluid we use the ratio of the asymptotic fluid speed to the asymptotic speed of sound, known as the asymptotic Mach number,
\begin{equation}
 \mathcal{M_{\infty}} = \frac{v_{\infty}}{c^{\infty}_{\rm{s}}}.
\end{equation}

In analogy to the relativistic Lorentz factor $W = \left(1-v^2\right)^{-\frac{1}{2}}$, we define the sound speed Lorentz factor $W_{\rm{c}} = \left(1-c_{\rm{s}}^2\right)^{-\frac{1}{2}}$ \citep{FI1}. Including these corrections, the relativistic Mach number is defined as \citep{Konigl:1980},
\begin{equation}
\mathcal{M}^{\rm{R}} = \frac{Wv}{W_{\rm{c}}c_{\rm{s}}}=\frac{W}{W_{\rm{c}}}\mathcal{M}\label{eq:rel_mach}.
\end{equation}
We further define the asymptotic relativistic Mach number as,
\begin{equation}
\mathcal{M}^{\rm{R}}_{\infty} = \frac{W_\infty v_\infty}{W_{\rm{c}}^{\infty}c^{\infty}_{\rm{s}}}=\frac{W_\infty}{W_{\rm{c}}^{\infty}}\mathcal{M}_{\infty}\label{eq:rel_mach2}.
\end{equation}
The relativistic Mach number is the ratio of the proper speed of the fluid with the proper speed of sound in the fluid, when this value is greater than one the flow is designated supersonic. When the Mach number is close to, but greater than one, we say that the flow is marginally supersonic.

The rest of the paper will proceed as follows; in section \ref{Sec:0} we present the coordinates used to study this problem. Section \ref{Sec:1}  discusses the equations of motion used in our ultrarelativistic system. Section \ref{Sec:2} presents the initial conditions and boundary conditions used to perform the simulations. In section \ref{Sec:3} we briefly cover the numerical methods developed for the simulations. Section \ref{Sec:4} is where we discuss the accretion profiles, while in section \ref{Sec:5} we discuss the accretion rates. In section~\ref{Sec:6} we present the flow morphology. Finally, section \ref{Sec:7} contains our conclusions.

In this paper we use geometric units, where $G=c=1$ with $c$ the speed of light in vacuum, and $G$ being Newton's gravitational constant. We also use the convention that Greek indices are spacetime indices, while Latin indices are purely spatial indices.

\section{Coordinates}\label{Sec:0}
As in our earlier study \citep{Penner:2011}, we are interested in the flow around a rotating black hole. The line element used is originally presented in \citet{PF1},
\begin{align}
ds^2=&-\left(1-\frac{2Mr}{\Delta}\right)dt^2+\frac{4Mr}{\Delta}dtdr\nonumber\\
&+\left(1+\frac{2Mr}{\Delta}\right)dr^2-2a\left(1+\frac{2Mr}{\Delta}\right)\sin^2{\theta}drd\phi\nonumber\\
&+\Delta d\theta^2 +-\frac{4aMr\sin^2{\theta}}{\Delta}dtd\phi\nonumber\\
&+(\Delta+a^2\left(1+\frac{2Mr}{\Delta}\right)\sin^2{\theta})\sin^2{\theta}d\phi^2,\label{eq:metric}
\end{align}
with
\begin{equation}
 \Delta = r^2+a^2\cos{\theta}^2.
\end{equation}
$a$ is the dimensionless measure of the rotation rate of the black hole and is related to the angular momentum of the black hole via $J=M^2a$, where $M$ is the mass of the black hole. For the present study we set $M=1$ without loss of generality.

\section{Equations of Motion}\label{Sec:1}

We obtain the equations of motion for the ideal ultrarelativistic hydrodynamic (UHD) system by using the conservation of stress-energy. The stress-energy tensor used to describe the UHD fluid is the same as that presented by \citet{Neilsen:2000}. To close the system of equations, we use an equation of state \eqref{eq:EoS} with $\Lambda=0$ to relate the internal energy density to the fluid pressure.
\begin{align}
\nabla_\mu T^{\mu\nu}&= 0
\end{align}
where $\nabla_\mu$ is the covariant derivative.
In the UHD limit we write the hydrodynamic stress-energy tensor:
\begin{equation}
T^{\mu\nu} = hu^\mu u^\nu +Pg^{\mu\nu}
\end{equation}
where $h$ is the enthalpy density of the system defined in \eqref{eq:enthalpy}.

We use the Anrowitt-Desner-Misner (ADM) $3+1$ formalism to re-express our system of partial differential equations \citep{York}, and use the Valencia formulation to describe our equations of motion. The equations of motion are presented in \citet{Neilsen:2000}. We adapt their equations to a stationary spacetime background as seen below.

The spatial momentum components are defined as,
\begin{equation}
 S_j = -n_\mu\gamma_{\nu j}T^{\mu\nu} = hW^2v_j,\label{eq:Momentum}
\end{equation}
and the energy,
\begin{equation}
E = n_\mu n_\nu T^{\mu\nu} = hW^2-P,\label{eq:Energy} 
\end{equation}
where $W$ is the relativistic Lorentz factor, $v_j$ are the 3+1 3-velocities, and $P$ is the fluid pressure.

This allows us to write the equations of motion as,
\begin{align}
\frac{\partial}{\partial t} \sqrt{\gamma} S_j + \frac{\partial}{\partial x^j} \sqrt{-g}\left(S_i\left( v^i-\frac{\beta^i}{\alpha}\right)\right. &+\left.P\delta^j{}_i\right)\nonumber\\
&= \sqrt{-g}T^{\mu\nu}\Gamma_{\mu\nu i}\label{eq:con1}\\
\frac{\partial}{\partial t} \sqrt{\gamma} E + \frac{\partial}{\partial x^i} \sqrt{-g}\left(E\left( v^i-\frac{\beta^i}{\alpha}\right)\right.&+\left.Pv^i\right)\nonumber\\
 &= \sqrt{-g}T^{\mu\nu}\Gamma^{t}{}_{\mu\nu}.\label{eq:con2}
\end{align}
%
\section{Initialization and Boundary Conditions}\label{Sec:2}
We use the method described by \citet{FI1,FI2,Penner:2011} to initialize the hydrodynamic variables, with the exception of the baryon density $\rho_0$ which is neglected in our model. 

The domain of integration for this study is defined by $r_{\min} \le r \le r_{\max}$ and $0\le\theta\le\pi$. $r_{\min}$ is determined in such a way that it will always fall inside the event horizon. The maximum radial domain, $r_{\max}$, was set to be sufficiently far from the event horizon that it would be effectively considered infinity. If the radial domain is not set to be large enough, the bow shocks interfere with the outer, upstream, boundary conditions which quickly destroy the simulation.

While previous studies assumed that a domain size $r_{\max}=50M$ was sufficiently large for ultrarelativistic systems, we discovered that this is not true for our system. Using such a small radial domain resulted in a simulation that was marred with unphysical boundary effects. We investigated this dependence and found that the formation of a steady state accretion flow was highly dependent on the radial distance between the black hole and domain edge. In particular we found that as the fluid becomes stiffer the larger the domain of integration needed to be. In this study we used regular grid spacing for both the radial and polar coordinates. In future studies we will investigate the use of a geometrically spaced grid. The geometric grid spacing is expected to provide higher accuracy near the event horizon of the black hole, while remaining numerically stable.

The numerical treatment of the boundary conditions for this problem are the same as those presented in \citet{Penner:2011,FI1}.
%
\section{Numerical Methods}\label{Sec:3}
The equations of motion (\ref{eq:con1}) and (\ref{eq:con2}) take on a general form,
\begin{equation}\label{eq:conservative}
\frac{\partial}{\partial t}\sqrt{\gamma}\:{\bf{Q}}+\frac{\partial}{\partial x^i}\sqrt{-g}\:{\bf{F}}^i({\bf{Q}}) = \sqrt{-g}\:{\bf{S}},
\end{equation}
where ${\bf{Q}}$ are the conservative variables, ${\bf{F}}^i$ denotes the flux, and ${\bf{S}}$ are the geometric source terms. To solve this system, we modified our high resolution shock capturing code used for the GRMHD (general relativistic magnetohydrodynamic) study presented in \citet{Penner:2011}. We changed the conservative variables, the primitive variable recovery scheme, and introduced a floor, as discussed below. Convergence tests may be found in App.~\ref{Appendix:1}.

Parallelization was performed using the Parallel Adaptive Mesh Refinement (PAMR). The PAMR infrastructure was developed by \citet{PAMR} and was built on the message passing interface C software. For the numerical simulations presented for models U1--U4 we use a $400\times160$ grid. When we increased the domain size we adjusted the radial grid to maintain the same spatial resolution. Simulations were performed using the Iridis cluster at the University of Southampton, the woodhen cluster at Princeton University, USA, and the WestGrid cluster of Canada.

\subsection{Primitive Variable Recovery}
Since we use a different set of primitive and conservative variables we describe the primitive variable recovery scheme. Our method used a modified version of Del Zanna's one parameter inversion scheme \citep{DelZanna:2003} most recently investigated by \citet{Noble}.

\noindent We define,
\begin{equation}
\Omega \equiv h W^2.\label{eq:Omega}
\end{equation}

The first step of the primitive variable recovery is to solve the conservative equations for the term $\Omega$, as outlined below.

\noindent We write the energy equation (\ref{eq:Energy}),
\begin{equation}
  E = \Omega - P,\label{eq:Newt_E}
\end{equation}
and using the momentum equation (\ref{eq:Momentum}) we get
\begin{equation}
 |S|^2(\Omega) = \gamma^{ij}S_iS_j = \Omega^2\left(1-\frac{1}{W^2}\right).\label{eq:Newt_S}
\end{equation}
We solve (\ref{eq:Newt_S}) for $W$,
\begin{equation}
 W(\Omega) = \left[ 1 -\frac{|S|^2}{\Omega^2} \right]^{-\frac{1}{2}},\label{eq:Newt_W}
\end{equation}
then re-formulate the equation of state, $P=(\Gamma-1)\rho$ as
\begin{equation}
 P(\Omega) = \frac{\Omega(\Gamma-1)}{\Gamma W^2}.\label{eq:Newt_Pressure}
\end{equation}
With the primitive variables defined as functions of $\Omega$ we use an iterative scheme to determine the values of the primitive variables.
Using Eqn.~\eqref{eq:Newt_E}, we define $f(\Omega)$,
\begin{equation}
f(\Omega) = \Omega - P - E= 0.\label{eq:NewtNewt}
\end{equation}
We then use Newton's method to solve the nonlinear equation (\ref{eq:NewtNewt}) for $\Omega$,
\begin{align}
\Omega^{I+1} &= \Omega^{I} - \frac{f(\Omega^I)}{
f^{\prime}(\Omega^I)},\\
f^{\prime}(\Omega^I) &= \frac{\partial f}{\partial\Omega}(\Omega^I),
\end{align}
where $I$ denotes the Newton iteration. To calculate $\partial f/\partial\Omega$, we use the relations listed below,
\begin{align}
\frac{d f}{d \Omega} &= 1 - P^{\prime}\\
\frac{dP}{d\Omega} &=
\frac{\left(W-2\Omega W^{\prime}\right)\left(\Gamma-1\right)}{\Gamma
W^3}\\
\frac{dW}{d\Omega} &=
-W^3\frac{|S|^2}{2\Omega^3},
\end{align}
where the prime denotes a derivative with respect to $\Omega$.
We perform this procedure at every time step, $n$. The Newton iteration must be initialized, so as an initial guess for the value of $\Omega^0$ we use the definition of $\Omega$ in Eqn.~\eqref{eq:Omega} and values for $h$, and $W$ from the previous timestep, $n-1$,
\begin{equation}
\Omega^0 = h^{n-1}/(1-\gamma_{ij}v_i^{n-1}v_j^{n-1})^2.
\end{equation}

Given $\Omega$, we solve for the primitive variables. Using \eqref{eq:Momentum} and \eqref{eq:Omega} we express the velocity components as;
\begin{equation}
 v_i = \frac{1}{\Omega}S_i.
\end{equation}
We then calculate the Lorentz factor $W$ and pressure $P$ from \eqref{eq:Newt_W} and \eqref{eq:Newt_Pressure} respectively.

\subsection{Floor}
There are instances where the code update routines produce conservative variables, where there are not any solutions for the primitive variables. This problem is independent of the inversion scheme used. In these instances, the primitive variables are reconstructed using interpolated values as determined by the primitive variables in surrounding cells. It is important to note that these instances were isolated to one or two cells in each time step, and also were only found in the downstream tail region of the evolution. Any effects due to the averaging were found to travel out of the domain, away from the black hole. It should also be noted that these numerical complications were restricted to the most extreme flow simulations.

\begin{table}
\begin{center}
\begin{tabular}{|c|c|c|c|c|}
\hline
Model & $\Gamma$ & $v_\infty$ & $R_{\rm{max}}$ & $M^{\rm{R}}_{\infty}$\\
\hline
U1  & $4/3$ & 0.3 & 100 & 0.44 \\
U2  & -     & 0.6 &  -  & 1.05 \\
U3  & -     & 0.7 &  -  & 1.38 \\
U4  & -     & 0.9 &  -  & 2.91 \\
U5  & $3/2$ & 0.6 & 200 & 0.75 \\
U6  & -     & 0.7 &  -  & 0.98 \\
U7  & -     & 0.71&  -  & 1.01 \\
U8  & -     & 0.9 &  -  & 2.06 \\
U9  & $1.1$ & 0.3 & 400 & 0.94 \\
U10 & -     & 0.32 &  - & 1.03 \\
U11 & -     & 0.6 &  -  & 2.27 \\
U12 & $5/4$ & 0.4 & 400 & 0.86 \\
U13 & -     & 0.5 &  -  & 1.02 \\
U14 & -     & 0.7 &  -  & 3.62 \\
\hline
\end{tabular}
\end{center}
\caption{Table of parameters used for the axisymmetric systems studied in this paper. We have selected the velocities to capture the flows in the subsonic, marginally supersonic, and the supersonic regime. We used the same velocity parameters for different values of the rotation parameter, $a=0.0,0.5,0.9,0.99$.}
\label{table:1}
\end{table}
%
%
\section{Accretion Profiles}\label{Sec:4}
In this section we describe the new and major features of each flow studied. This is done by presenting cross sections of the pressure accretion profiles found in Table \ref{table:1}.

\begin{figure*}
\begin{minipage}{\textwidth}
\centering
 \includegraphics[width=3in]{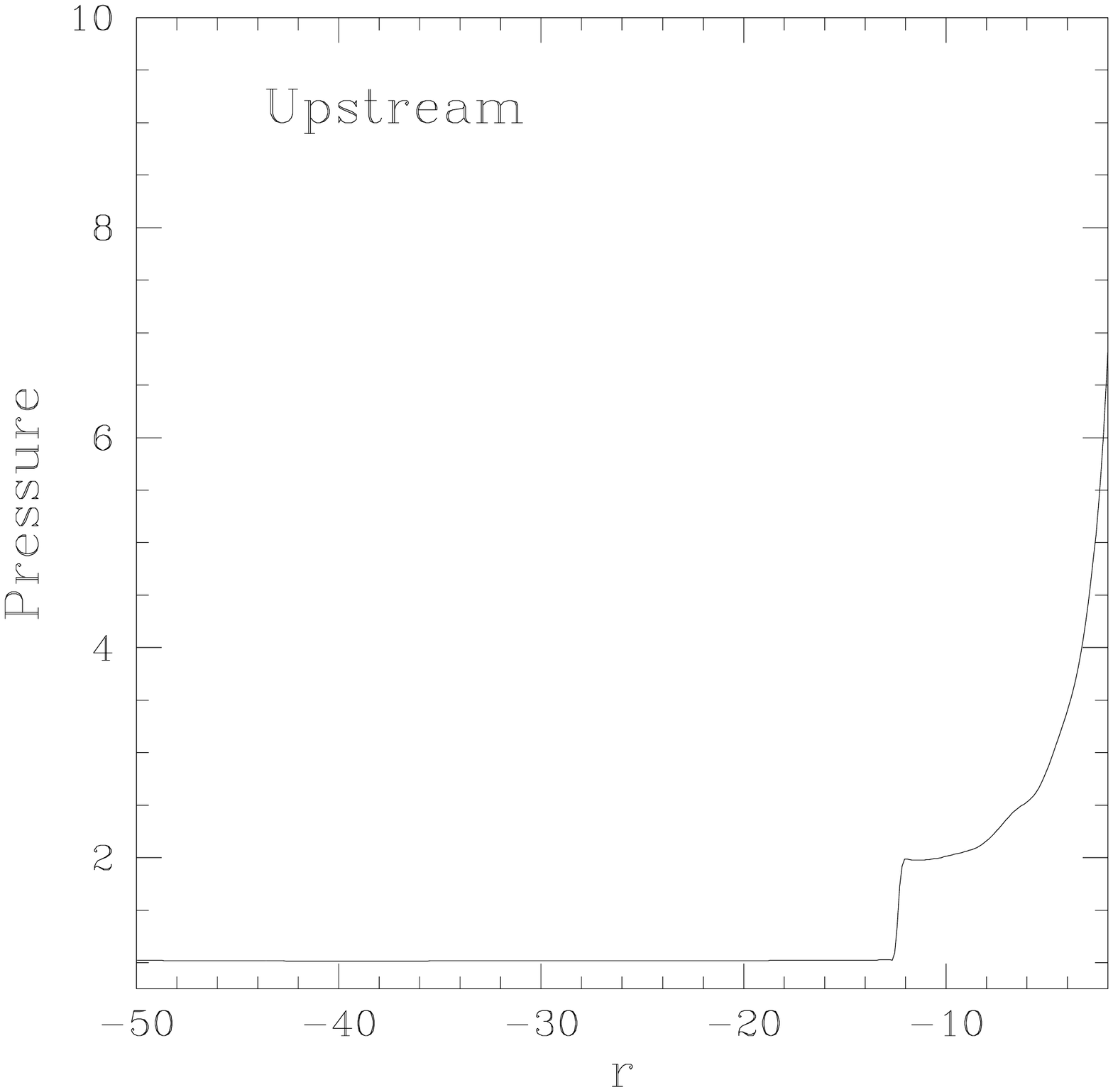}
 \includegraphics[width=3in]{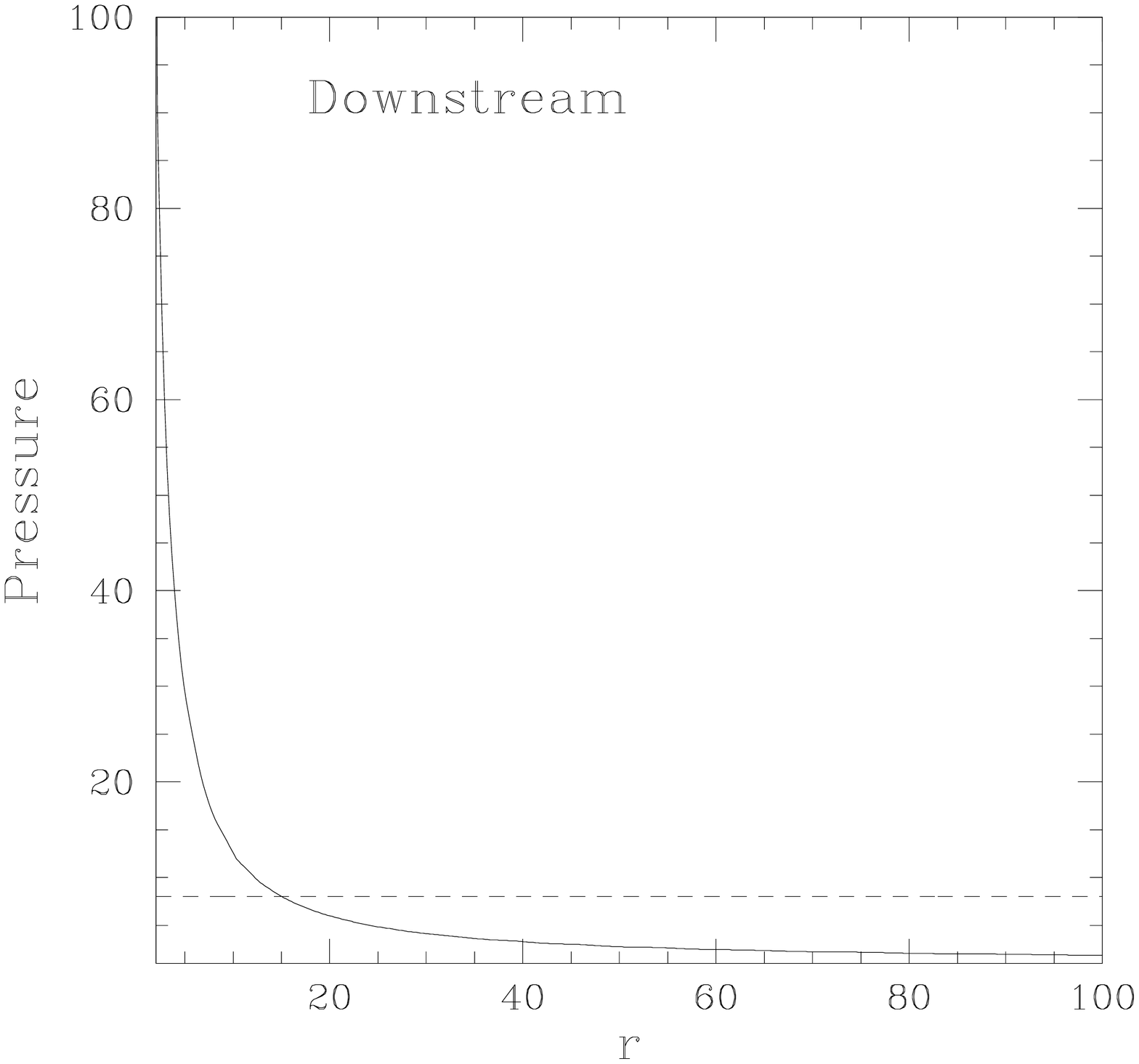}
\caption{Ultrarelativistic pressure profile upstream (top) and downstream (bottom) for model U2 with radial domain $r_{\rm{EH}}\le r\le 100$. To emphasize the bow shock in the upstream region we restrict our attention to the region $-50 \leq r \leq -r_{\rm{EH}}$. The pressure in the upstream region of the black hole is small by comparison to the pressure in the downstream region, or wake. The pressure profile in the upstream region indicates that there is a shock in the upstream region of the black hole. The above profiles were taken at $t=1000M$. With the larger radial domain, we see that the pressure profile smoothly matches the upstream boundary conditions. The dashed line in the downstream profile emphasizes the maximum pressure of the fluid surrounding the black hole in the upstream side.}\label{fig:axi6}
\end{minipage}
\end{figure*}

\begin{figure*}
\centering
\begin{minipage}{6.5in}
  \includegraphics[width=3in]{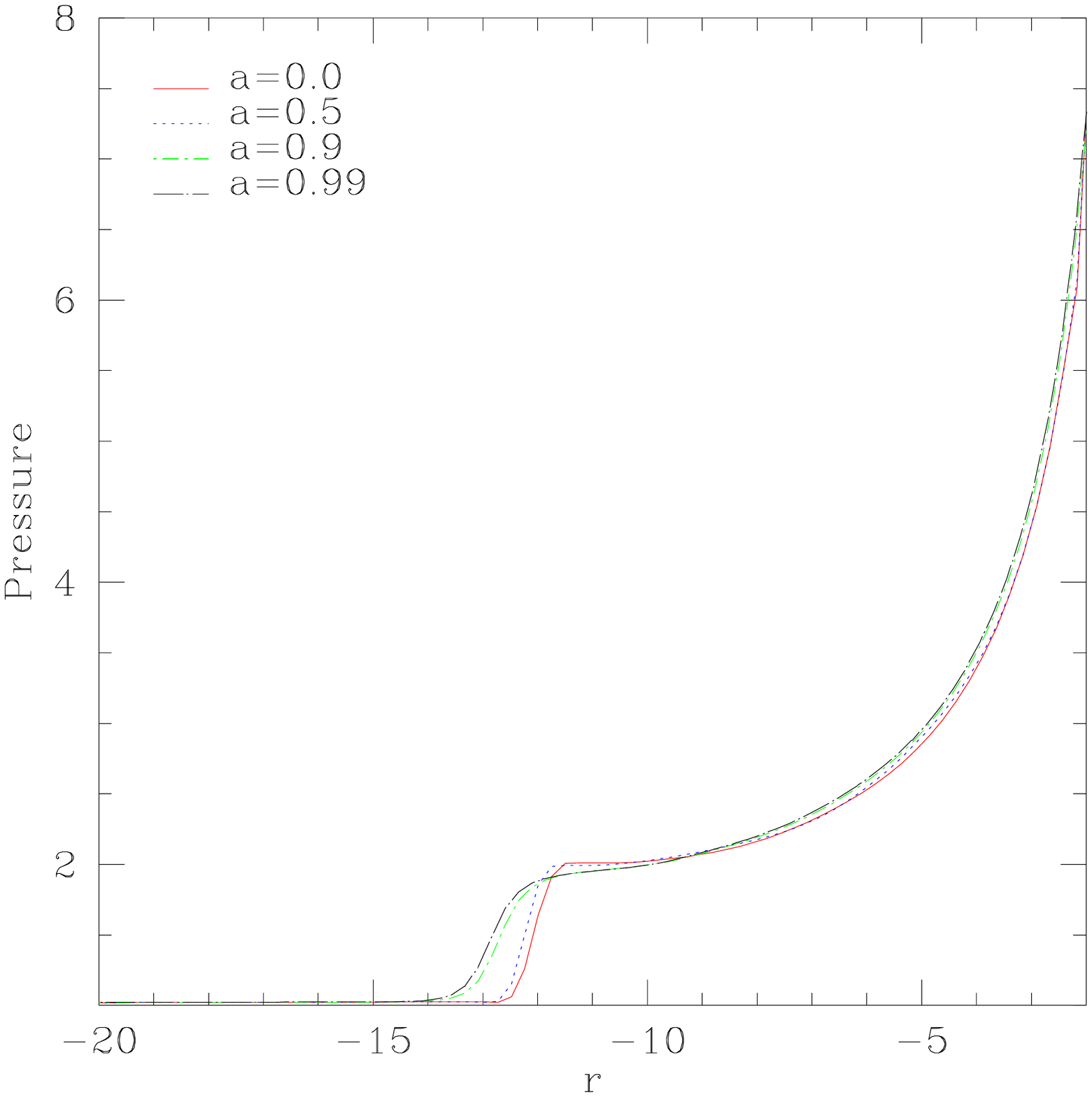}
  \includegraphics[width=3in]{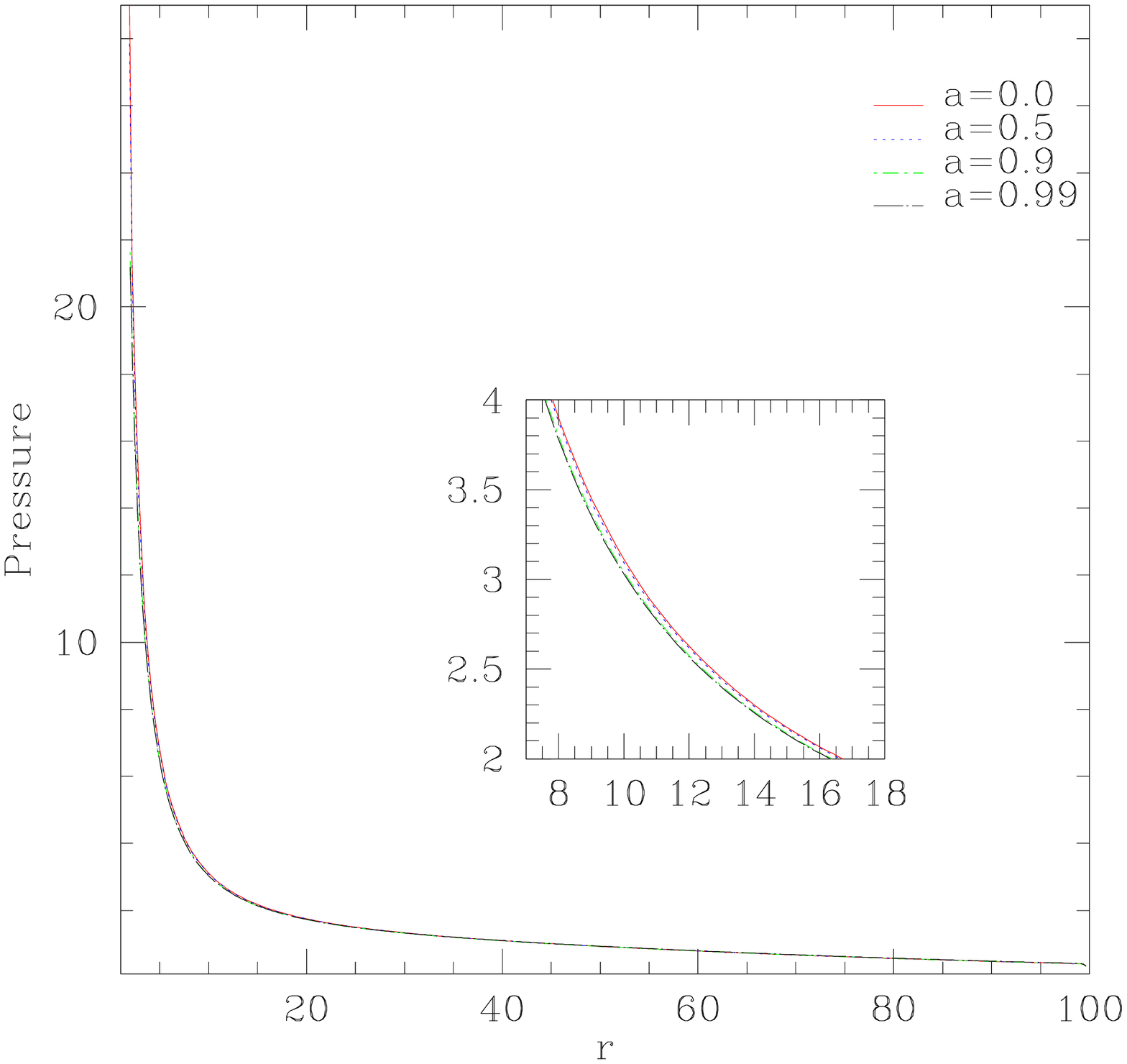}
\end{minipage}
\caption{Ultrarelativistic pressure profiles for the rotating axisymmetric accretion problem for model U2, with $a=0.0, 0.5, 0.9, 0.99$. We see that when the black hole rotates the bow shock persists. The upstream plot focuses on the region $r_{\rm{EH}}\le r \le 20$ to emphasize the bow shock region. We see the effects of rotation on the sharpness of the bow shock. As the spin of the black hole increases the resulting shock front begins to smooth. Since the shock front for $a=0.9$ is only marginally sharper than that for $a=0.99$, we suspect that the spin rate would need to exceed the maximal value allowed \emph{i.e.}~$a>1$ to prevent the existence of a bow shock. Recall that if $a>1$ we have a naked singularity. }\label{fig:Pressure_v_6_all}
\end{figure*}

The first profiles we investigate are those for models U1--U4 with $a=0$. We found, for the supersonic black holes with $v_{\infty}=0.6$ and $v_{\infty}=0.9$, that the pressure profiles along the axis of symmetry were dramatically different. In Fig.~\ref{fig:axi6} we present the pressure profile for model U2 where we see a bow shock has formed. As we expect from previous hydrodynamic studies \citep{FI1}, the upstream pressure profile shows that the maximum upstream pressure is much smaller than the maximum pressure on the downstream side of the black hole along the event horizon. The unexpected feature of this flow is the presence of the bow shock, which was not obtained in previous studies due to the parameter range investigated. The detached or bow shock seen in the left plot in Fig.~\ref{fig:axi6} appears to be a feature of the marginally supersonic flows in UHD systems. The detached shock persists when the black hole rotates, which may be seen in Fig.~\ref{fig:Pressure_v_6_all}.

\begin{figure*}
\centering
\begin{minipage}{\textwidth}
  \includegraphics[width=3in]{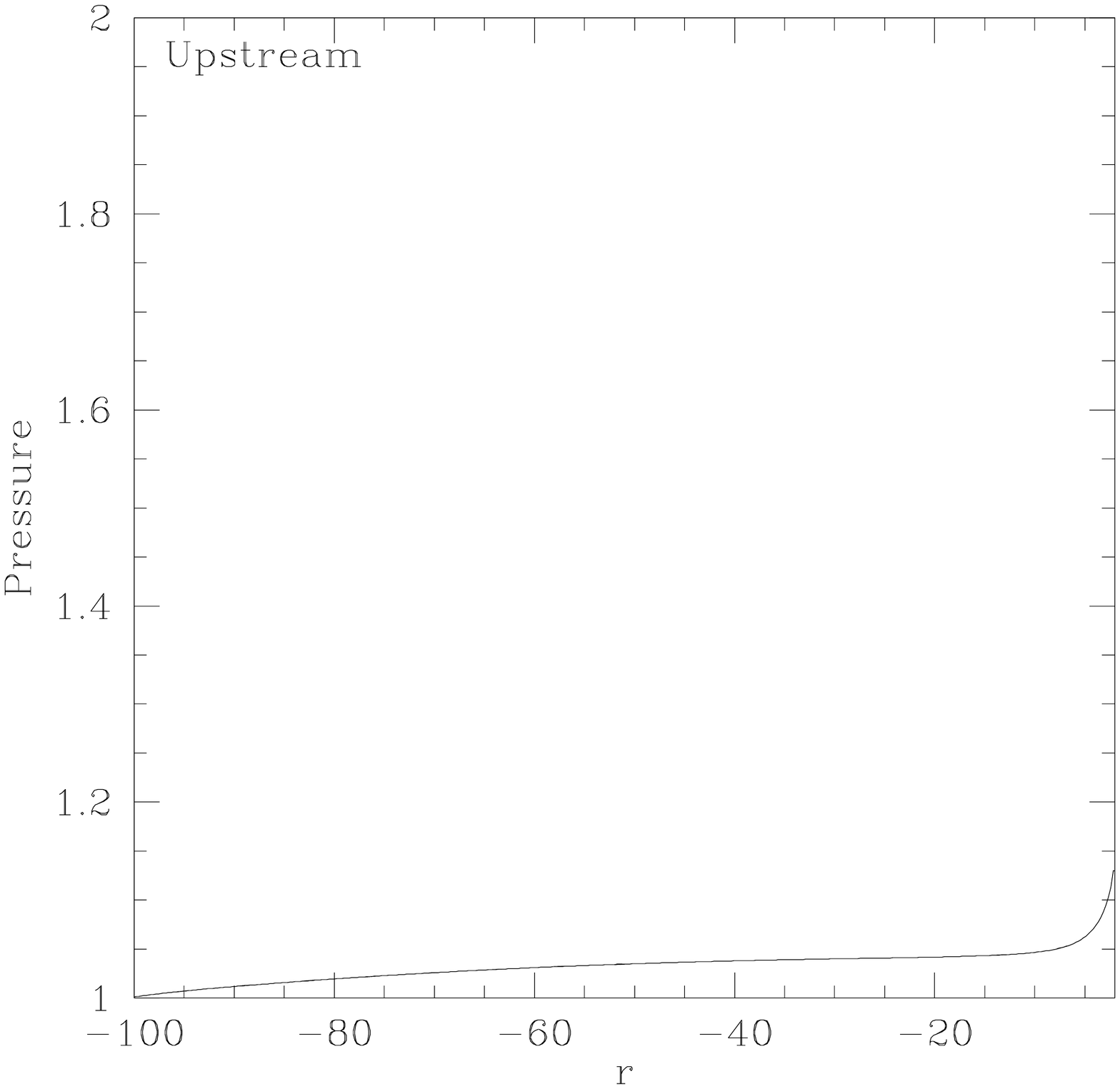}
  \includegraphics[width=3in]{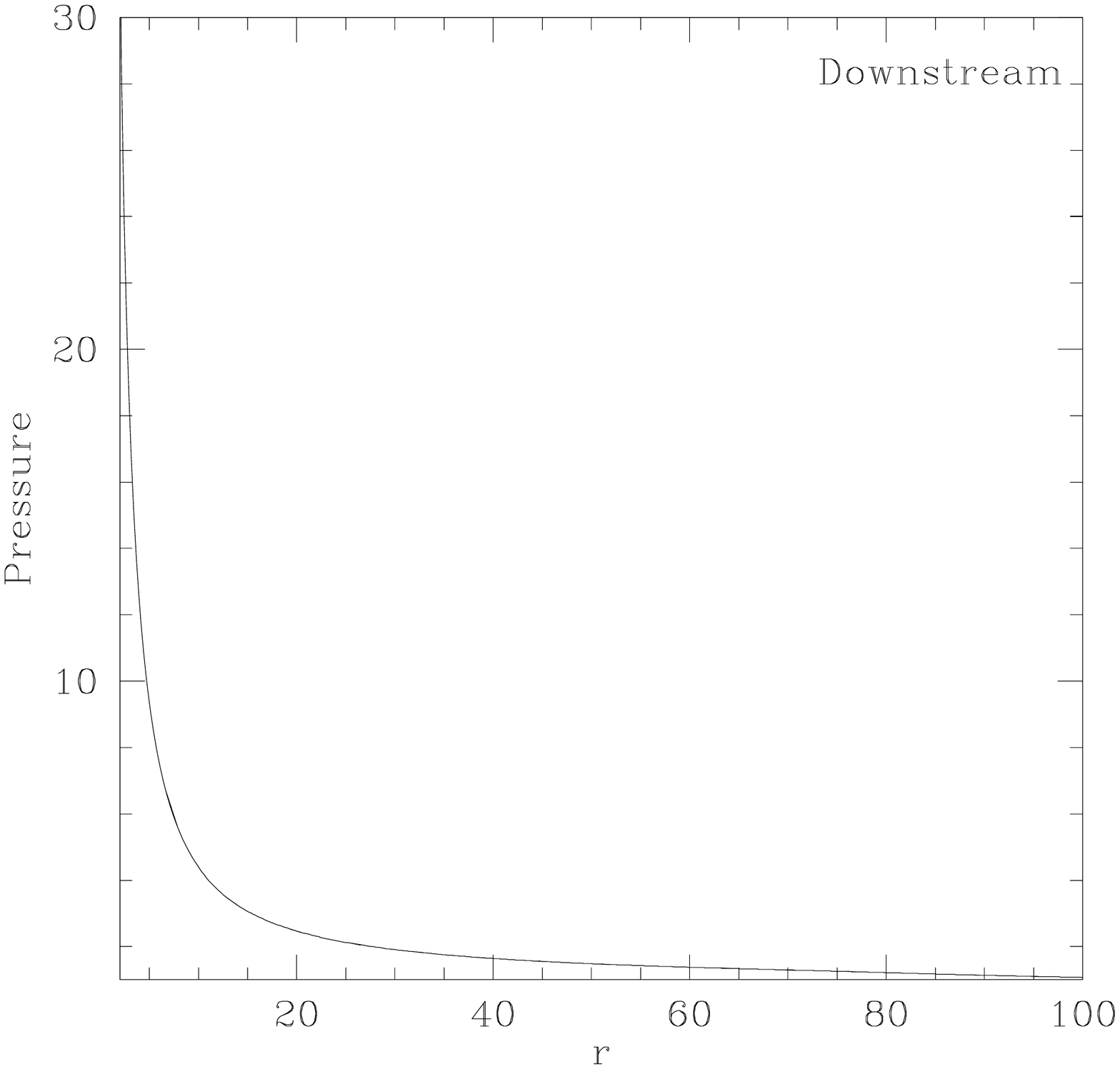}
\caption{In the left plot we show the pressure cross-section in the upstream region along the axis of symmetry plotted from $r_{\rm{EH}}\le r \le r_{\max}=50$ for model U3. In the 
right plot we observe the pressure cross-section in the downstream region along the axis of symmetry. We see that the pressure is maximal on $r_{\rm{EH}}$, the radial location of the event horizon.}\label{fig:axi7}
\end{minipage}
\end{figure*}

\begin{figure*}
\begin{minipage}{\textwidth}
\centering
  \includegraphics[width=3in]{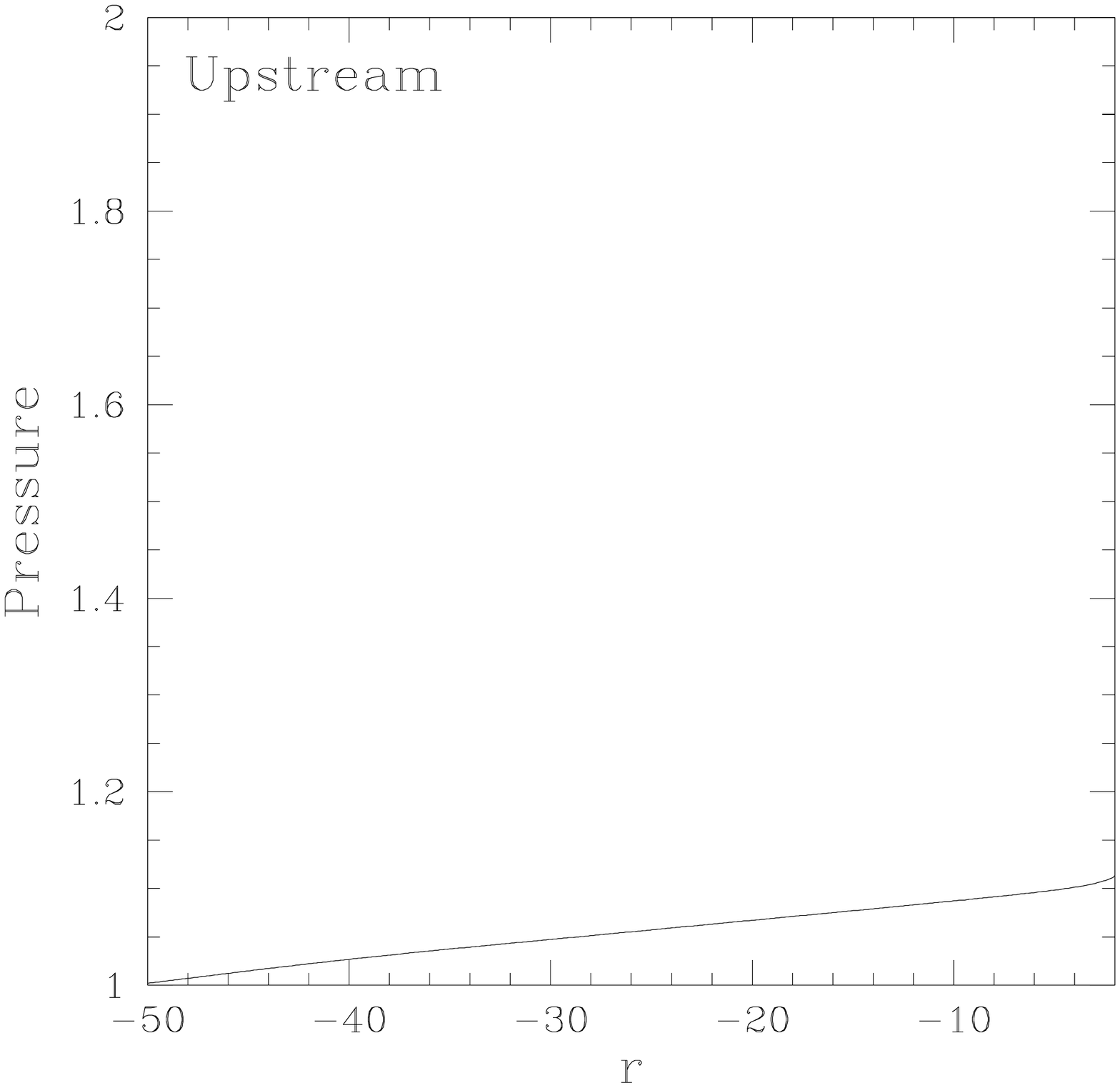}
  \includegraphics[width=3in]{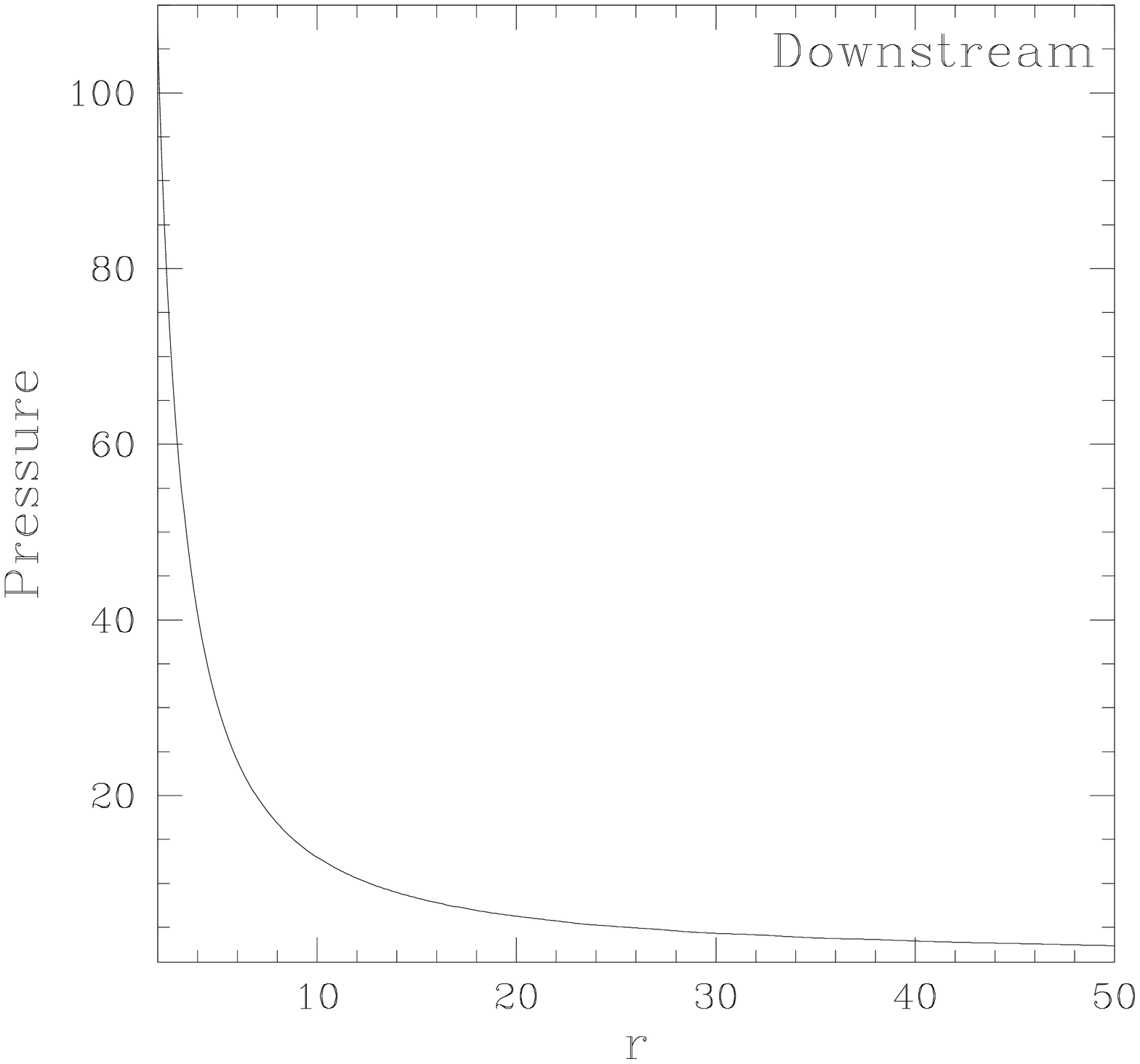}
\caption{The pressure cross-section in the upstream region along the axis of symmetry plotted for model U4 in the region $r_{\rm{EH}}\le r \le 50$. As with model U3 in Fig.~\ref{fig:axi7} the pressure is maximal on $r_{\rm{EH}}$. The downstream pressure on the event horizon dwarfs the upstream pressure on the event horizon by two orders of magnitude.}\label{fig:axi9}
\end{minipage}
\end{figure*}

When we increase the flow of the fluid past the black hole we find that the detached shock is prevented, as seen in Figs. \ref{fig:axi7} and \ref{fig:axi9}. Model U3 begins to show signs that the tail shock will detach from the black hole; however, the shock stabilises near the front of the black hole, never making contact with the axis of symmetry. We see a snapshot of the stable tail shock in Fig.~\ref{fig:P_43_v_all}.

When our simulation begins the fluid passes the black hole and focuses on the downstream side of the black hole, just as seen in previous studies \citep{FI1}. As the simulation proceeds the tail shock that forms widens, increasing the Mach cone opening angle. The fluid continues to flow past the black hole interacting with the pressure from fluid in the tail. The tail shock causes a back-pressure in the oncoming fluid. Depending on the value of the asymptotic velocity one of two things happen. If the velocity is high enough, as seen in models U3 and U4 in the bottom panels of Fig.~\ref{fig:P_43_v_all}, the pressure in the tail shock will balance the fluid pressure from the oncoming fluid. Consequently, the tail shock stabilises. If, on the other hand, the asymptotic fluid velocity is not sufficiently high as in models U1 and U2 in the top panels of Fig.~\ref{fig:P_43_v_all}, the pressure in the tail shock will continue to build, over-powering the oncoming fluid pressure. In this instance the tail shock will continue to widen, the angle of attachment will move to the upstream side of the black hole, make contact with the axis of symmetry, and detach from the black hole. We see evidence of this when comparing models U2 and U3. While a bow shock does not form in U3, we see that the tail shock attaches to the black hole in the upstream region. The pressure from the oncoming fluid in model U3 is sufficient to balance the tail shock, and thus the tail shock persists.

When we investigate black holes with more extreme asymptotic speeds such as seen in model U4 we recover the traditional tail shock seen in previous hydrodynamic studies \citep{FI1}. We present a snapshot of model U4 for all four spin parameters investigated in Fig.~\ref{fig:P_43_v_all}. It is clear that the tail shock persists.

As we would expect for systems with tail shocks, when looking at the on-axis pressure profiles for models U3 and U4 in Figs.~\ref{fig:axi7} and \ref{fig:axi9} respectively, the upstream pressure is dwarfed by the downstream pressure on the event horizon. By using horizon penetrating coordinates, we are sure this is not a coordinate effect as is cautioned by \citet{PF1}.

When we investigated model U4 we obtain the expected tail shock presented in similar hydrodynamic studies \citep{FI1}.

\begin{figure*}
\centering
\begin{minipage}{6.5in}
  \includegraphics[width=3in]{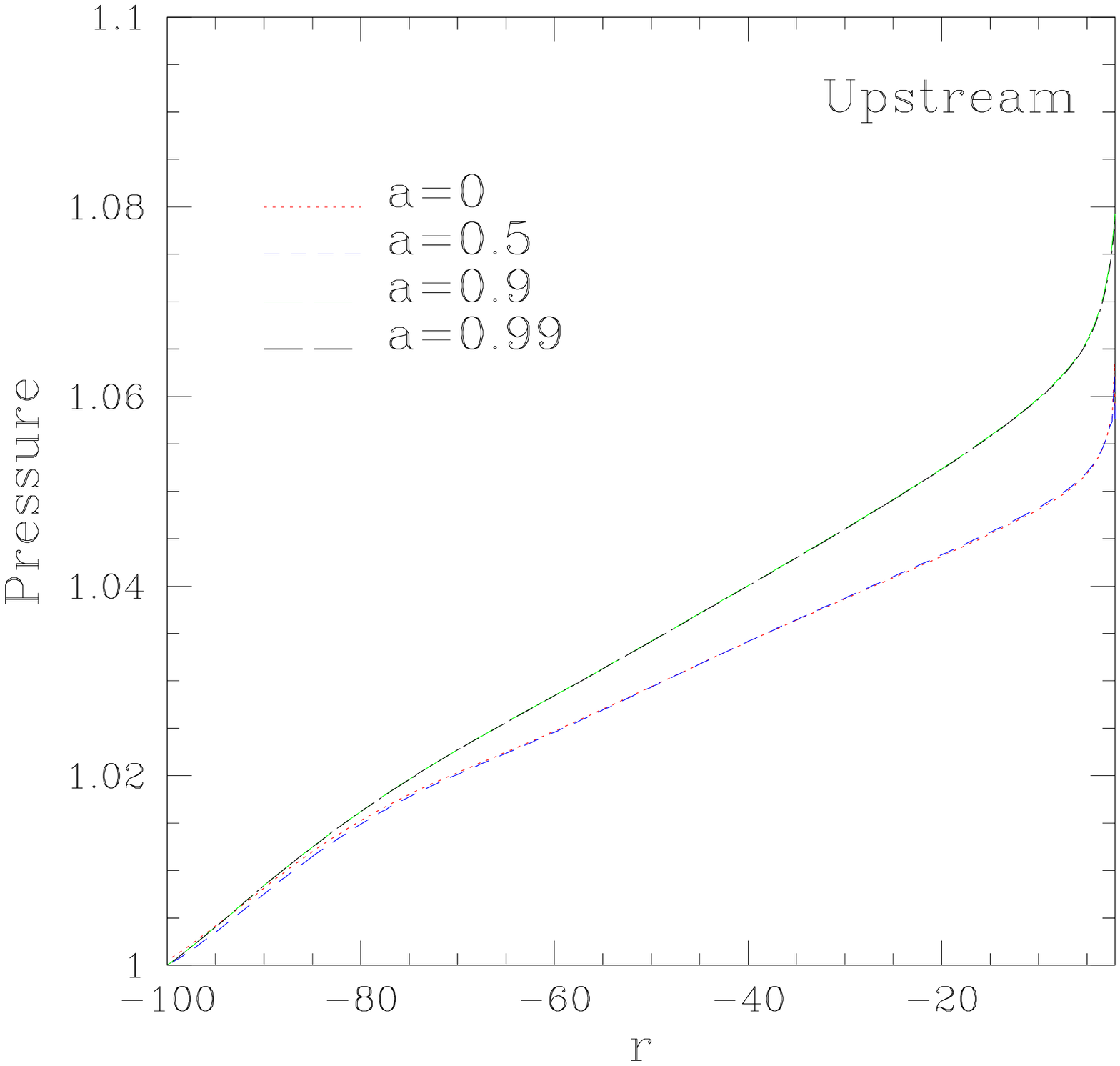}
  \includegraphics[width=3in]{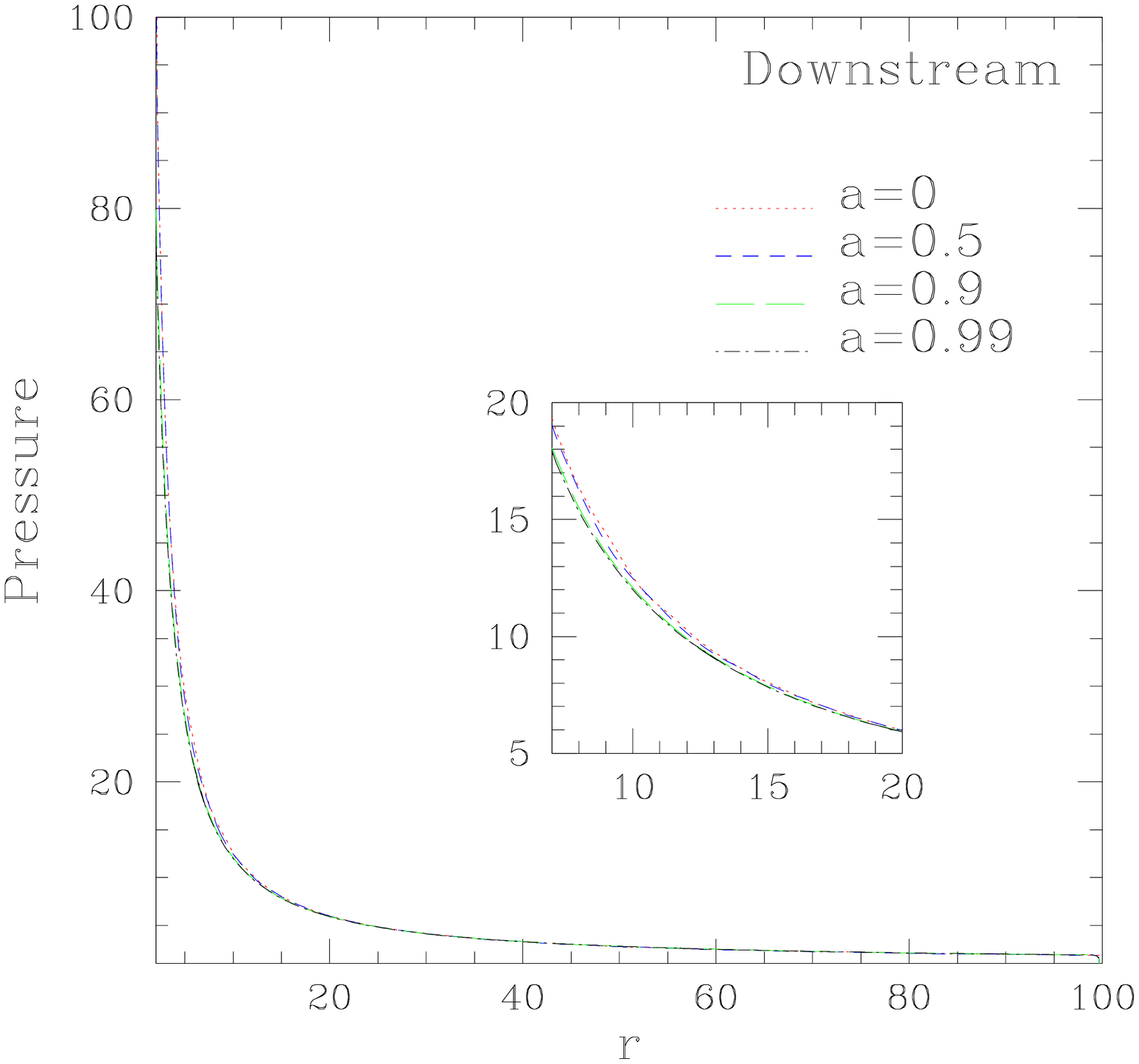}
\end{minipage}
\caption{Ultrarelativistic pressure profiles for the rotating axisymmetric accretion problem for model U4, with $a=0.0, 0.5, 0.9$. We see that for black hole spin parameters $a=0$ and $a=0.5$ the downstream (right) pressure profiles are only marginally distinguishable. As the spin is increased to $a=0.9$, we see a much larger pressure in the upstream region of the black hole (left). The profiles on the downstream region become slightly sharper when the spin parameter is increased.}\label{fig:Pressure_v_9_all}
\end{figure*}

\begin{figure}
\centering
 \includegraphics[width=3in]{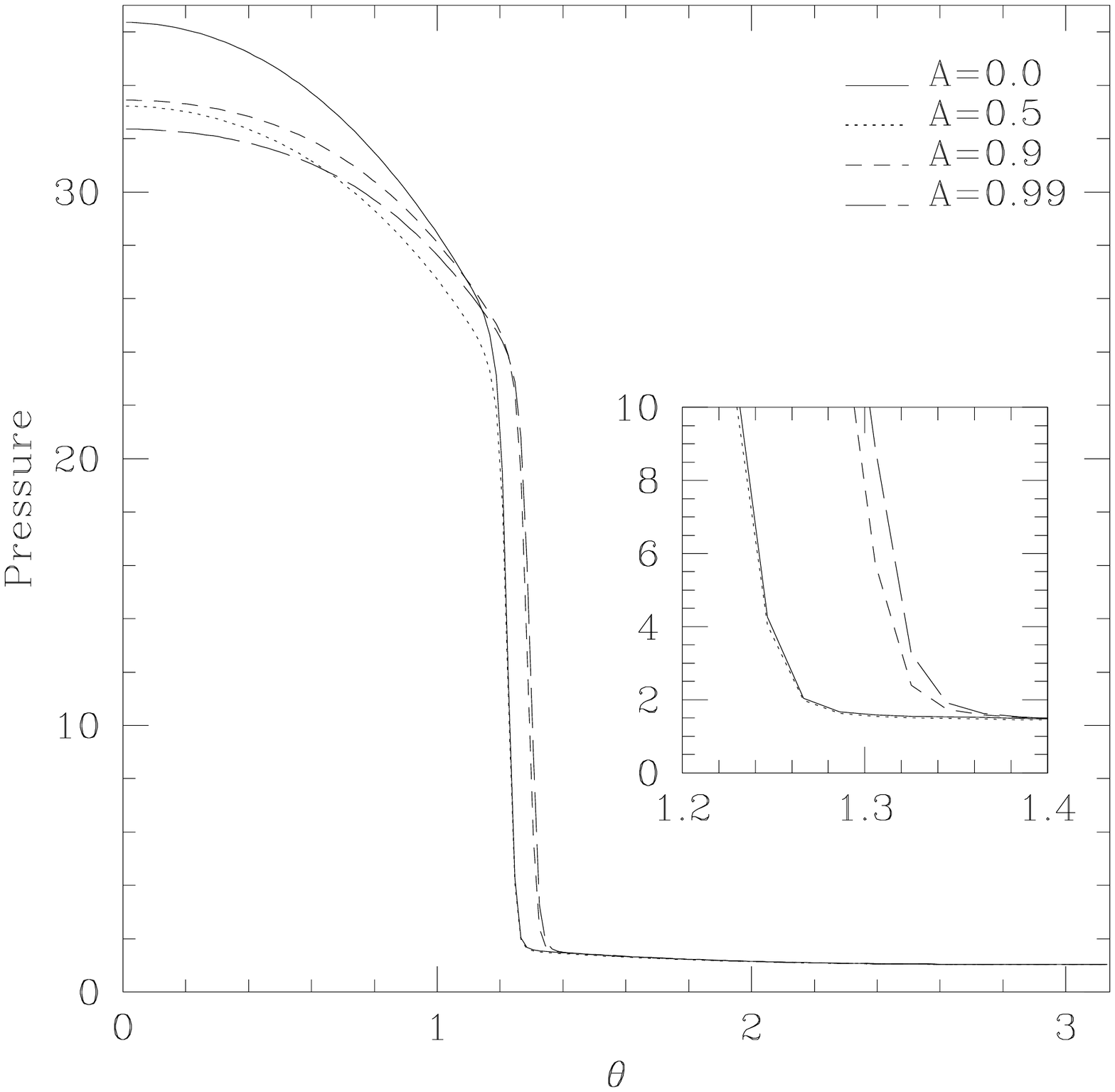}
\caption{Ultrarelativistic axisymmetric pressure profile $P(1000,4,\theta)$ for a section through the tail shock at $r=4\rm{M}$ for model U4 with spin parameters $a=0.0,0.5,0.9$. The shock opening angle for black holes with spin greater than $a=0.5$ is larger than that for smaller spin rates.}\label{fig:Pressure_cut_v_9_all}
\end{figure}

We also investigated axisymmetric accretion onto a rotating black hole with the axis of rotation aligned with the axis of symmetry. The upstream pressure profiles were only marginally different for black holes with $a=0$ and $a=0.5$ as seen in the left plots in Fig.~\ref{fig:Pressure_v_9_all}. When the spin rate was increased to $a=0.9$ the upstream profile noticeably changed. For all spins investigated the upstream pressure remained low relative to the maximum downstream pressure. The downstream pressure profiles themselves were not appreciably altered, as is seen in the inset plot of Fig.~\ref{fig:Pressure_v_9_all}. 
For a comparison of the effects of the black hole spin on the accretion profile we refer the reader to Fig.~\ref{fig:spin_43} where we present a snapshot of model U4 for all four spin parameters investigated. The spin rate of the black hole did not have a significant impact on the presence of a tail shock.
The effect of the rotation of the black hole is more readily apparent in the angular cross section seen in Fig.~\ref{fig:Pressure_cut_v_9_all}. As the rotation rate of the black hole is increased the pressure on the axis of symmetry decreases while the opening angle of the Mach cone increases. However, for the black hole spin rate to impact the type of shock exhibited the asymptotic fluid velocity will have to be slower than those investigated in this study. For the given parameters the black hole spin parameter will most likely effect the fluid flows with $0.6 < v_\infty < 0.7$. We would also expect a parameter to exist that places the bow shock on the upstream side of the black hole.
When we investigate other adiabatic constants we see similar behaviour in all ultrarelativistic systems, as may be seen in \ref{Appendix:2}. Models of stiffer fluids have a faster speed of sound, and are expected to produce results with a wider Mach cone. Since our parameter survey searches for marginally supersonic parameters as well as subsonic, we find that if we do not use a sufficiently large domain of integration that the tail shock will come in contact with the upstream outer boundary. This causes an unphysical pressure build-up at the outer domain that quickly dominates the fluid simulation. Thus for slower fluid flows we found that the domain of integration had to increase to capture the dynamics we are interested in.

\section{Accretion Rates}\label{Sec:5}

For our ultrarelativistic system we measure the energy accretion rate,
\begin{align}
\dot{E} &= \oint_{\partial V}{\sqrt{-g}T^{tr}d\phi} + \int_{V}{T^{\mu\nu}\Gamma^t{}_{\mu\nu}drd\phi},
\end{align}
where $g$ is the determinant of the metric \eqref{eq:metric}.

Since we are simulating an ultrarelativistic fluid we cannot directly calculate the baryon mass accretion rate as performed in \citet{FI1,FI2,Penner:2011}.
 
\begin{figure}
\centering
  \includegraphics[width=3in]{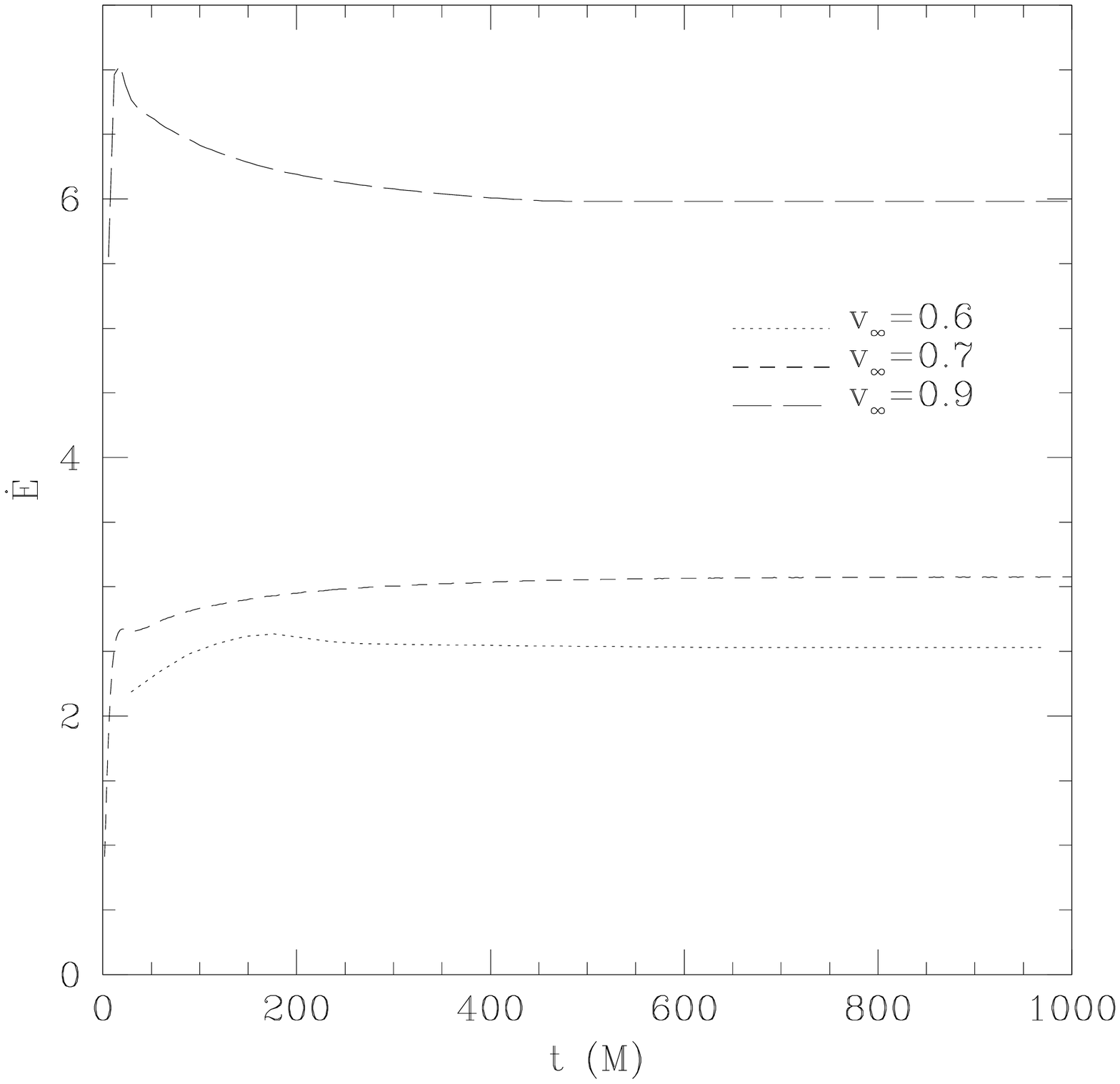}
\caption{The energy accretion rates for models U2 to U4 with $a=0$. We see that in all cases a steady state solution is obtained. Note that the amount of energy accreted by the central object is related to the velocity of the fluid relative to the central object.}\label{fig:energy_43}
\end{figure}

\begin{figure*}
\centering
  \includegraphics[width=3in]{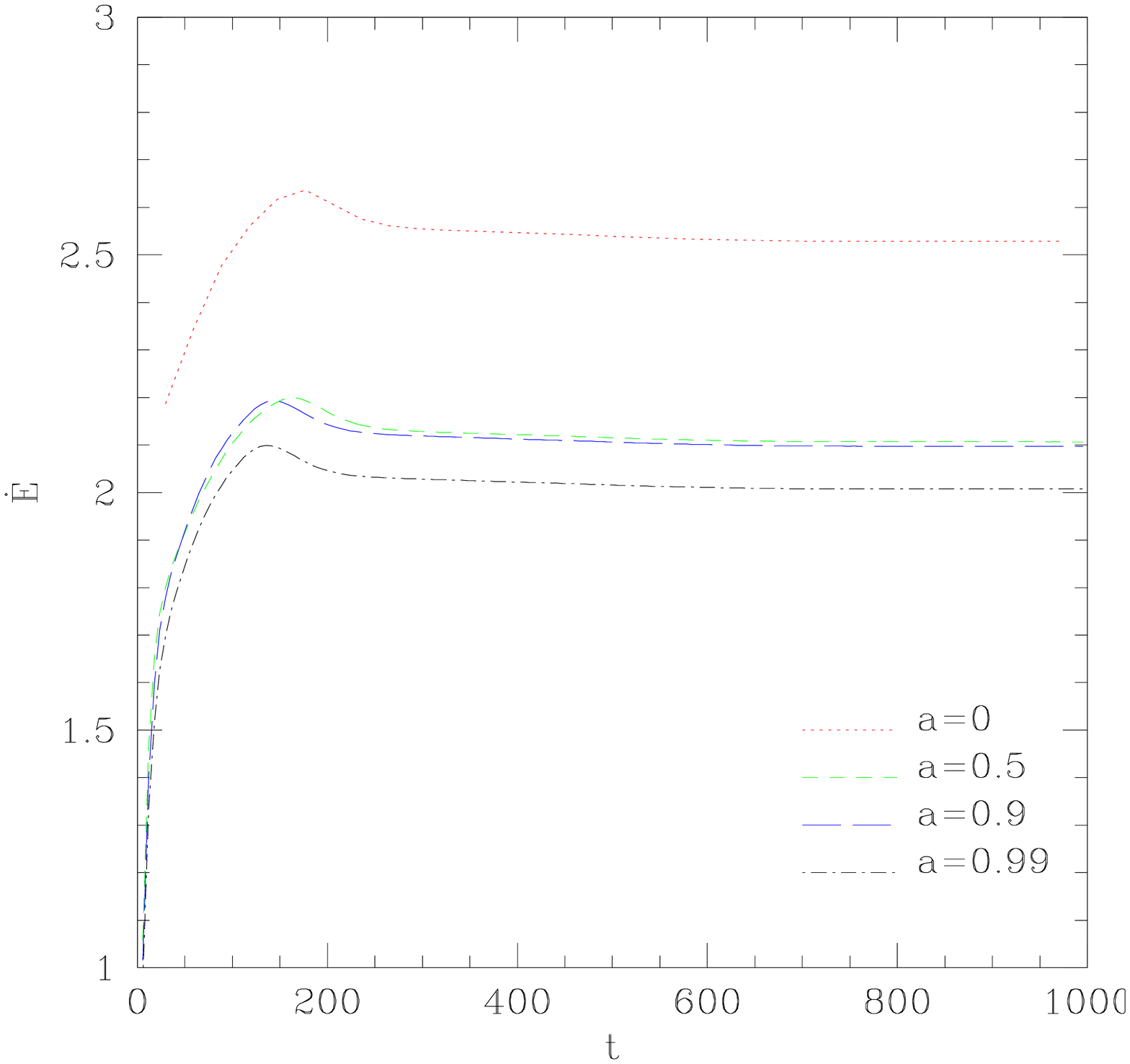}
  \includegraphics[width=3in]{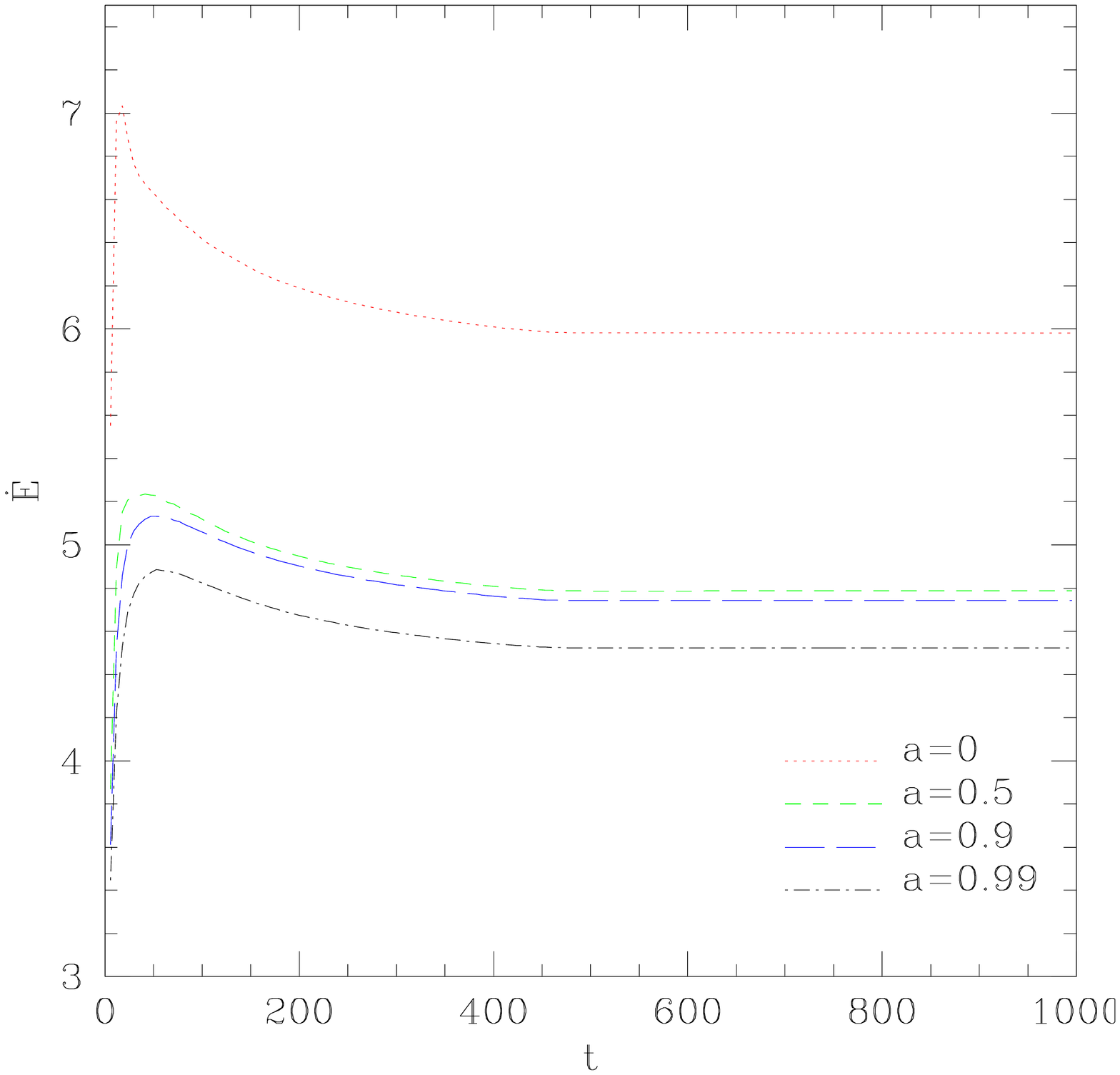}
\caption{The energy accretion rates for models U2 (left) and U4 (right) for $a=0,0.5,0.9,0.99$. We see that the energy accreted by the black hole depends on the spin parameter $a$. As the spin parameter is increased the energy accretion drops significantly. The energy accretion rate is measured over a surface defined by the event horizon. As the spin parameter increases this surface decreases, thus we expect a decrease in the measured energy flux.}\label{fig:energy_a}
\end{figure*}

As in our previous study we measure the accretion rates at the Schwarzschild radius. We rescale the accretion rates with the equivalent Bondi accretion rates as determined from our simulations when the black hole is stationary relative to the asymptotic background fluid. The energy accretion rates may be seen in Figs.~\ref{fig:energy_43} and \ref{fig:energy_a}.

We see in figure \ref{fig:energy_a} that the models with slower asymptotic speeds (models U1,U2) reach a steady state around $t=200M$ and for the faster asymptotic speeds (models U3, U4) the steady state is reached near $t=600M$. The plots for the remaining models may be found in App.~\ref{Appendix:2}. We note, when comparing Fig.~\ref{fig:energy_11} or Fig.~\ref{fig:energy_54} with Fig.~\ref{fig:energy_32}, that for supersonic flows with a lower the adiabatic constant, the faster the system reaches the final steady state.
%
\section{Flow Morphology}\label{Sec:6}

\begin{figure*}
\begin{minipage}{\textwidth}
\centering
  \includegraphics[width=2in]{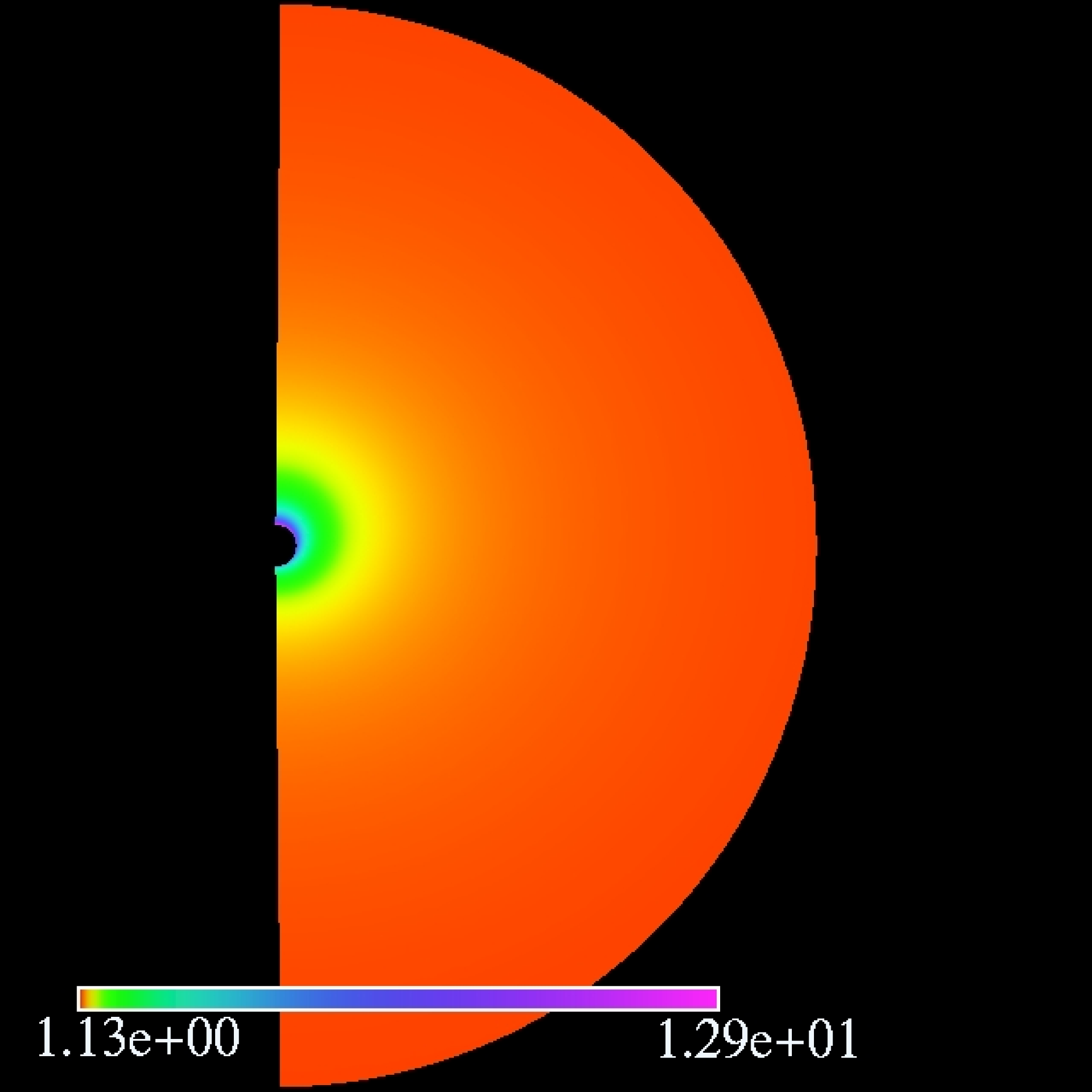}
  \includegraphics[width=2in]{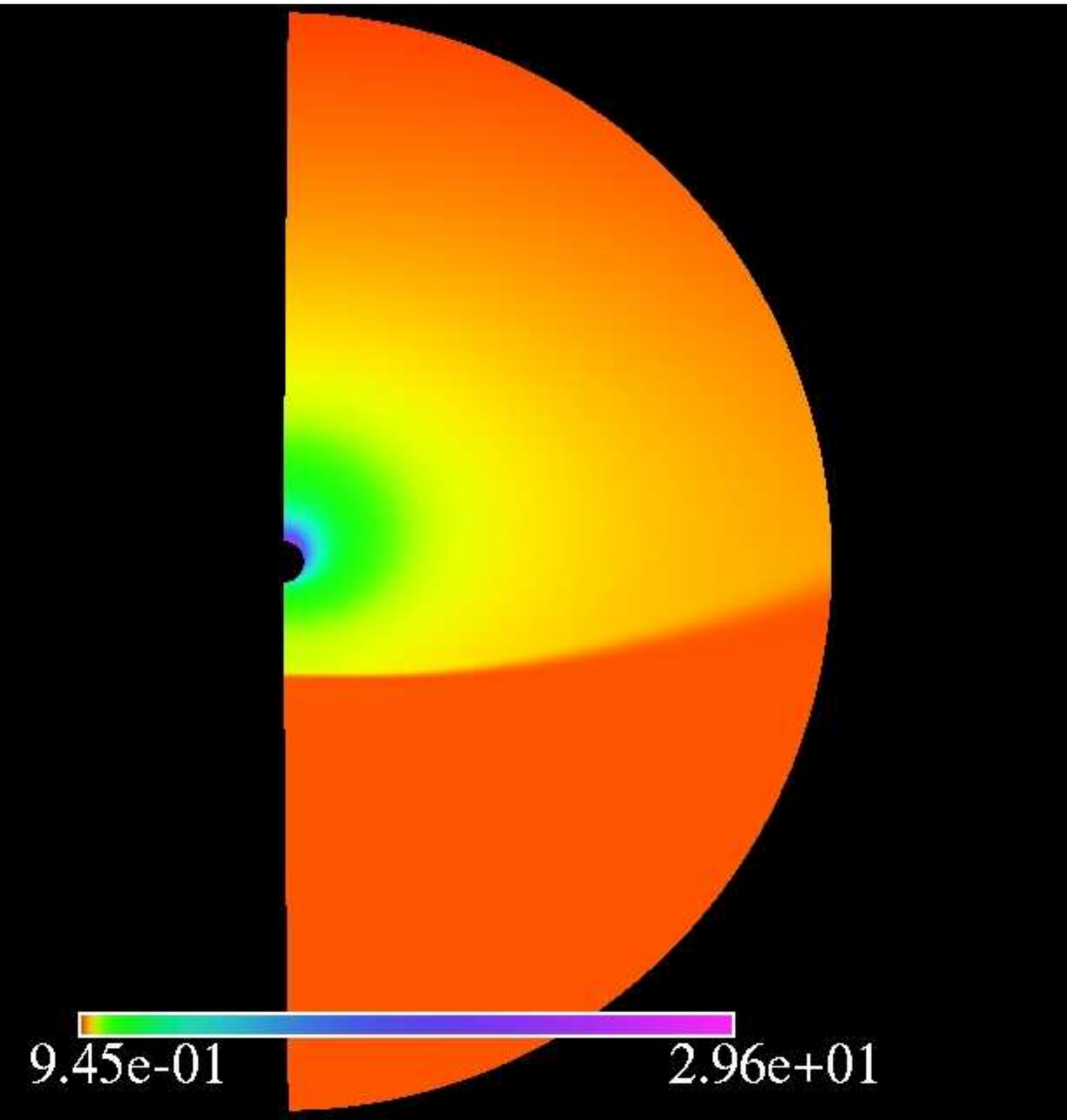}\\
  \includegraphics[width=2in]{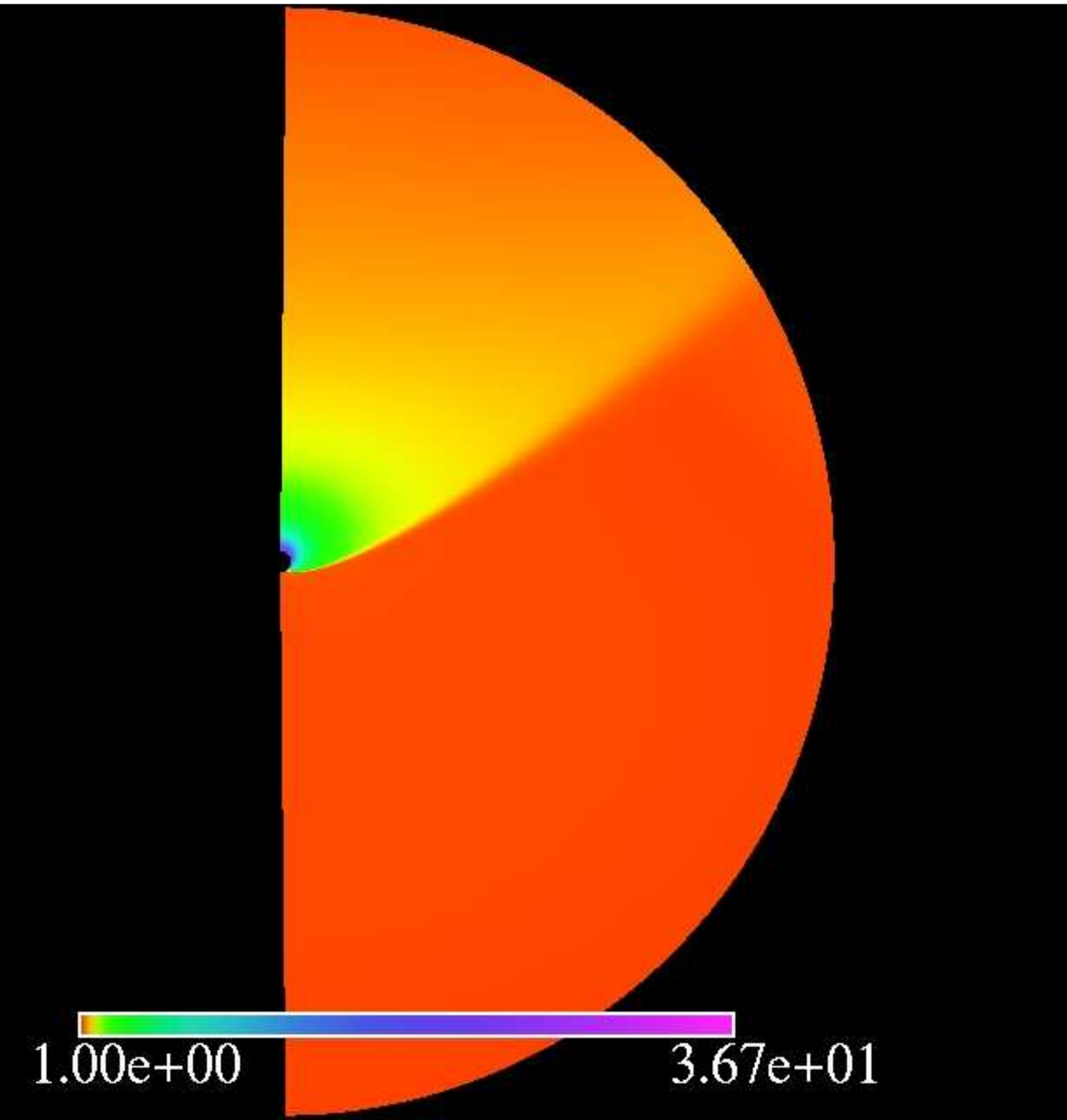}
  \includegraphics[width=2in]{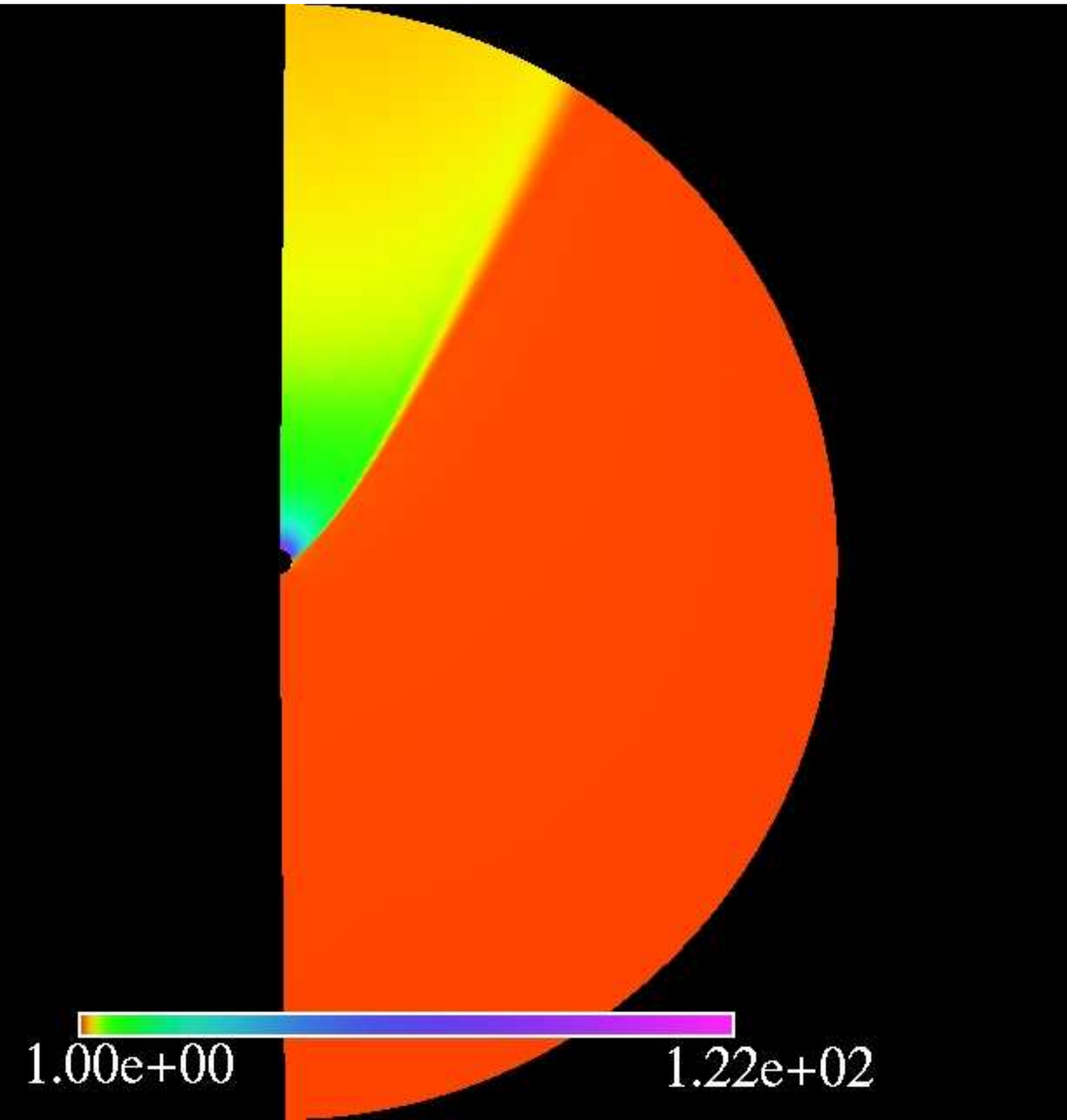}
\caption{Pressure profile for the spherically symmetric evolution for models U1 with $\mathcal{M}^{\rm{R}}_{\infty}=0.44$ (top left), U2 with $\mathcal{M}^{\rm{R}}_{\infty}=1.05$ (top right), U3 with $\mathcal{M}^{\rm{R}}_{\infty}=1.38$ (bottom left), and U4 with $\mathcal{M}^{\rm{R}}_{\infty}=2.91$ (bottom right), when a steady state is achieved. Model U1 was terminated at $t=400M$ due to boundary effects disrupting the simulation. We see that the black hole in model U4, traveling well above supersonic speeds, produces a readily apparent tail shock; however, when the flow is only marginally supersonic, as seen in model U2, we see the presence of a bow shock. If the flow is increased, as in model U3, we see that the tail shock persists, indicating there is some critical value where the bow shock remains fixed in the upstream region of the black hole along the event horizon. In the Newtonian studies, the presence of the bow shock was essential to the formation of what is known as the flip-flop instability. In our studies, the accretion rate on the event horizon remains constant and visualization of the flow remains constant, thus we conclude that the flow is steady despite the presence of a bow shock. In the first panel the flow is subsonic and so consequently, takes on a profile similar to that of Bondi accretion.}\label{fig:P_43_v_all}
\end{minipage}
\end{figure*}

The results of the simulations using the parameters found in table \ref{table:1} are discussed here. We present the final state of model U1--U4 for both the spherically symmetric $(a=0)$ and axisymmetric $(a\ne0)$ black holes. Each of the parameters studied here establish a steady state solution. As in the general hydrodynamic models \citep{FI1}, we find that in supersonic flows, the upstream region of the flow is smooth, while in the downstream region there is the presence of a Mach cone. This is seen in the bottom right panel of Fig.~\ref{fig:P_43_v_all}. While for marginally supersonic flows such as seen in the bottom left panel of Fig.~\ref{fig:P_43_v_all}, the shock attaches to the black hole in the upstream region. In both models U3 and U4, the attached shock extends downstream, to the outer domain edge.

\begin{figure*}
\begin{minipage}{6in}
\centering
  \includegraphics[width=2in]{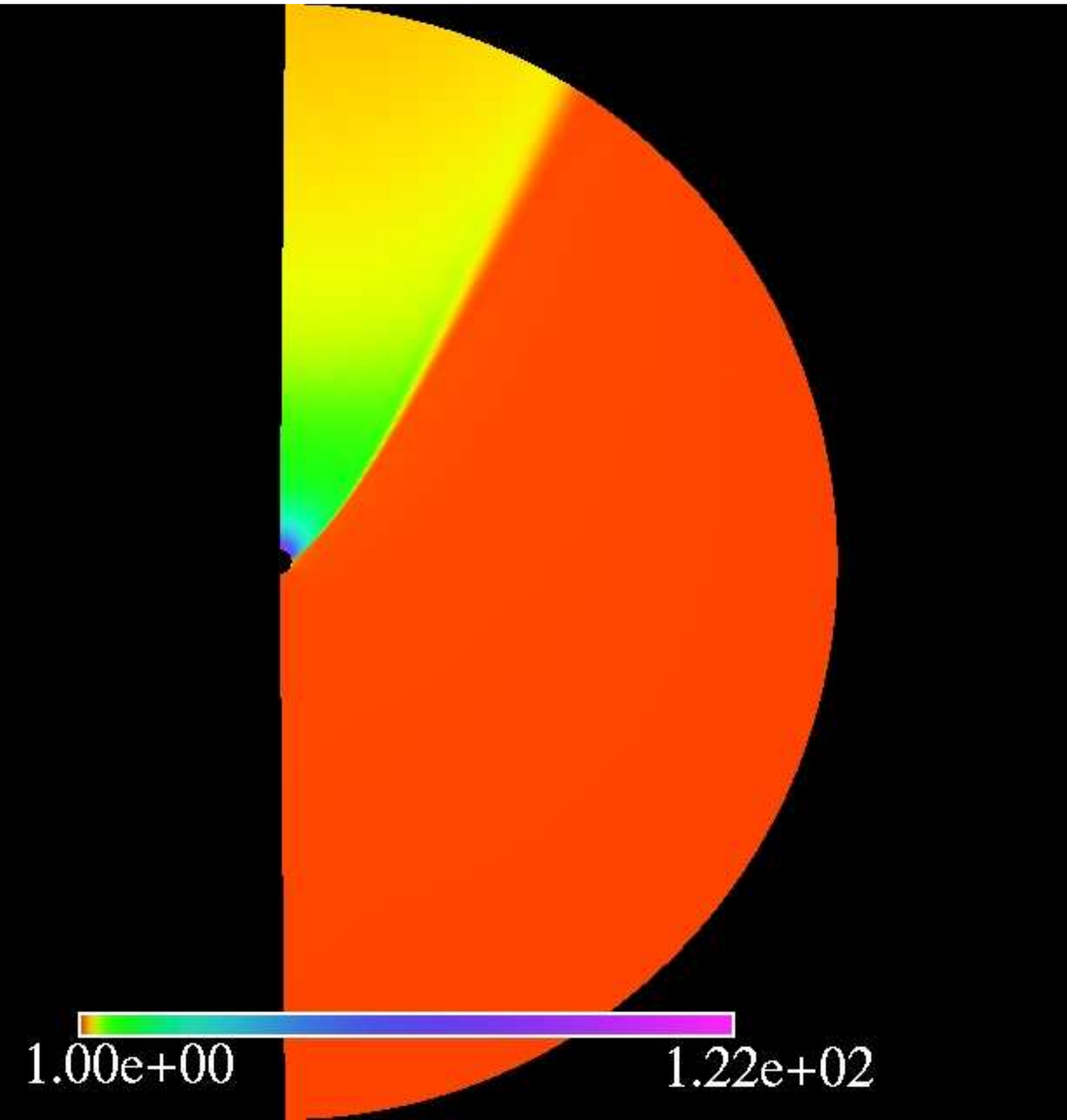}
  \includegraphics[width=2in]{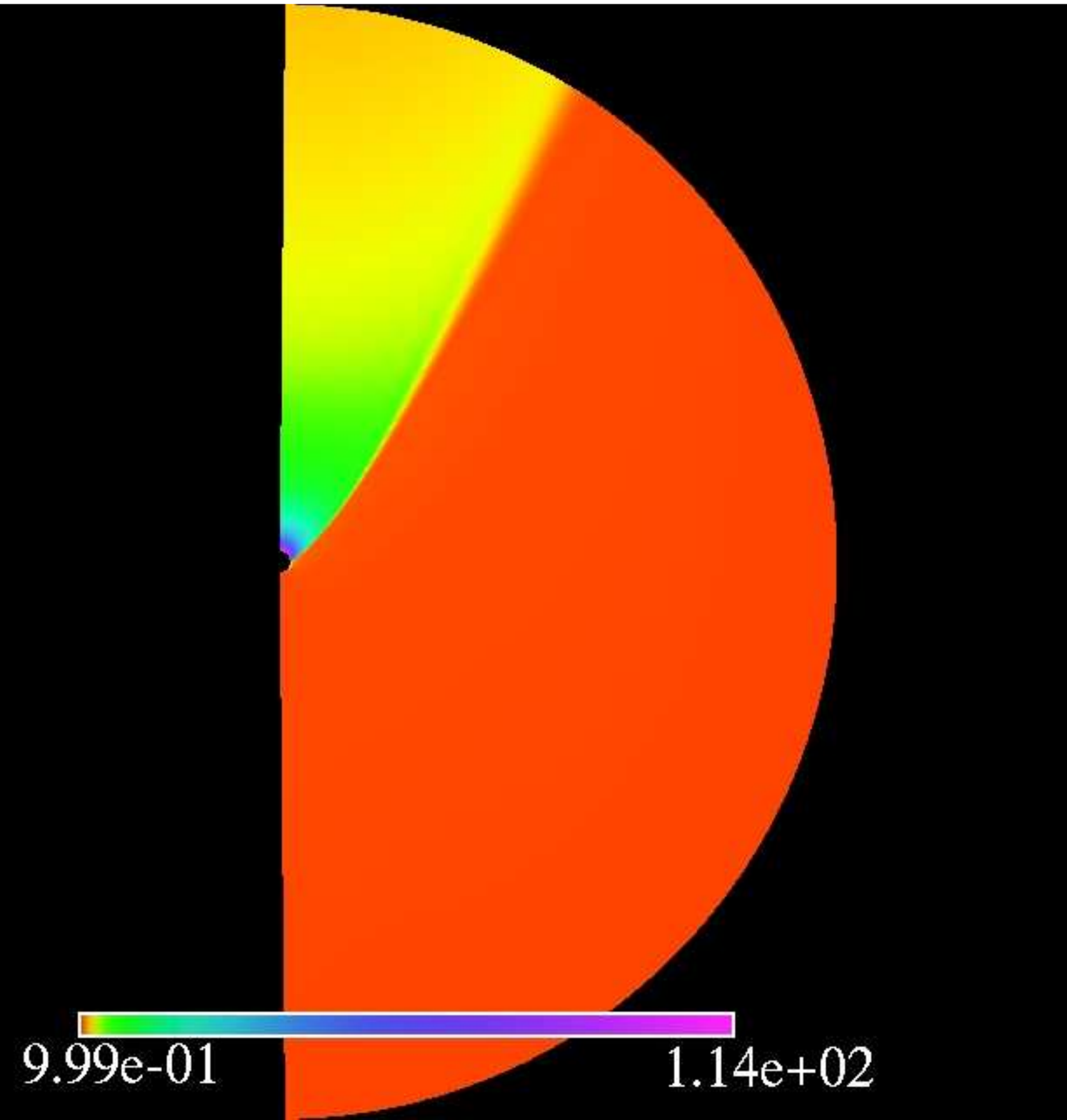}\\
  \includegraphics[width=2in]{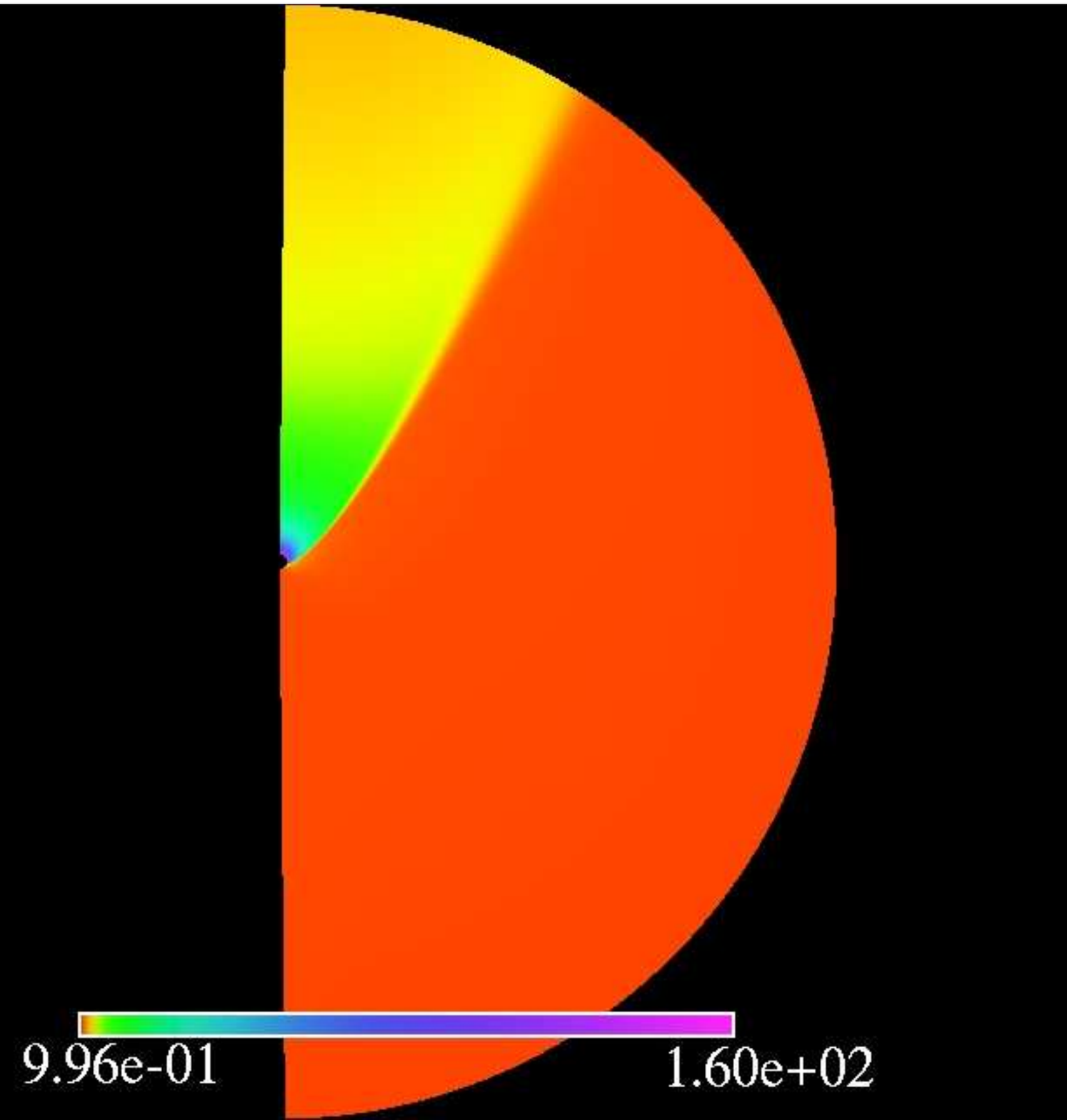}
  \includegraphics[width=2in]{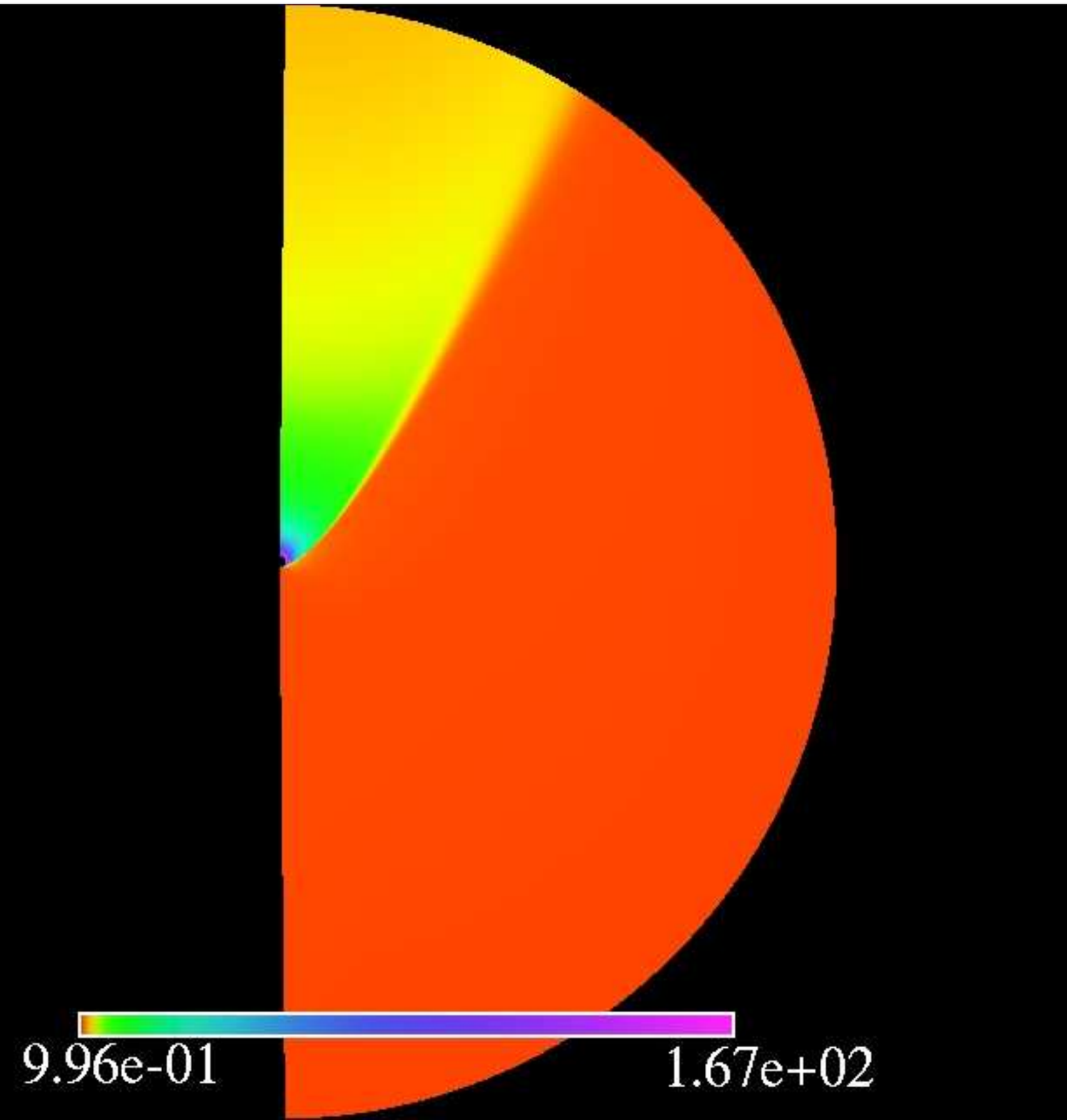}
\caption{Pressure profile for the $a=0.0$ (top left), $a=0.5$ (top right), $a=0.9$ (bottom left) and $a=0.99$ (bottom right) axisymmetric evolution for model U4 at $t=400M$. We see that a black hole traveling well above supersonic speeds produces a readily apparent tail shock. As the spin parameter of the black hole is increased we see that the point of contact for the tail with the event horizon of the black hole migrates upstream in agreement with the results presented in Fig.~\ref{fig:Pressure_cut_v_9_all}. All of these flows reach a steady state. Performing the same simulation using negative rotation rates produced identical pressure profiles and energy accretion rates.}\label{fig:spin_43}
\end{minipage}
\end{figure*}

The top two panels in Fig.~\ref{fig:P_43_v_all} present the flow patterns during the steady state phase of the fluid flow for models U1 and U2 respectively. As expected for subsonic flows, seen in the top left panel, the pressure profile does not exhibit any shocks. The flow pattern changes when we increase the flow rate beyond the speed of sound. For $\Gamma=4/3$, $c_s\sim0.57$ so the asymptotic velocity $v_\infty=0.6$ used for model U2 is marginally supersonic. The top right panel of Fig.~\ref{fig:P_43_v_all}, exhibits a detached/bow shock. Even in the presence of a bow shock the flow retains its steady state behaviour. Models U2 and U3 appear to bound a special bow shock that persists along the event horizon.
As seen in Fig.~\ref{fig:spin_43}, when the black hole spins, with the spin axis aligned with the fluid axis of symmetry, the fluid profiles do not change appreciably relative to a non-rotating $(a=0)$ black hole. This behaviour is expected since a detailed inspection of the pressure cross sections in Fig.~\ref{fig:Pressure_cut_v_9_all} reveals that the cross-sections change only slightly as a function of spin. The upstream fluid experiences a slight increase in pressure while the downstream region remains similar for all spin parameters. Since these changes are only slight, we expect the overall fluid morphology to be similar, as seen in Fig.~\ref{fig:Pressure_cut_v_9_all}. The fact that the spin rate does not have a large impact on the fluid morphology is seen if we look back at the upstream pressure profiles for model U2 in Fig.~\ref{fig:Pressure_v_6_all}, where the bow shock persists with the spin turned on. We have seen that the shock opening angle is dependent on the black hole rotation rate, but for the parameters investigated, we also find that the rotation rate does not impact the type of shock that forms.
%
\section{Conclusions}\label{Sec:7}
We have shown that there is a relation between a system reaching a steady state and the size of the domain of integration. The lower the adiabatic constant the larger the domain needs to be prevent boundary effects from disrupting the simulation. In fact we also see that this is particularly important for subsonic flows which have detached shocks traveling upstream.

Our simulations did not indicate the presence of any instabilities. This is likely due to the reduction in the complexity of the overall system. In the traditional black hole accretion problems we are concerned with both the rest-mass interactions and the energy, while in our system the rest-mass contribution is negligible. Studies using axisymmetry tend to fail to capture instabilities due to the restrictive nature of the symmetry. This does not rule out the existence of instabilities similar to the standing accretion shock instability \citep{Fernandez:2010} or those presented by \citet{Donmez:2010}, where the instability timescale is very large, and consequently was not discovered using short-time simulations.

It may be argued that running our simulations could be run for longer times, it becomes questionable how well we can approximate the accreting body as maintaining a fixed mass-energy, or how long before issues such as dynamical friction become important. Certainly, we expect that the assumption that the background fluid has constant pressure will become unreasonable over sufficiently long timescales.

Our next study investigates the UHD system described above using an infinitely thin-disc accretion model, which relaxes the axisymmetry. According to the studies by \citet{Donmez:2010} using the usual hydrodynamic approximation, the thin-disc symmetry allows the formation of a flip-flop instability.

The ultrarelativistic flow past a spherically symmetric black hole is in strong agreement with the flow past an axisymmetric/rotating black hole with the axis of rotation aligned with the direction of the fluid flow. The morphology of both geometric configurations is very similar and does not deviate too far from the results of previous studies of the full hydrodynamic evolution. Ultimately, most parameters surveyed resulted in a steady state solution. The only exception has been the $\Gamma=1.1$ models which require a large amount of computational resources to reach a steady state. The accretion rates exhibited a dependence on the location of the computational boundary, such that a larger domain was essential to capture the proper flow patterns for some adiabatic constants. We are currently investigating the effects of magnetic fields on this same configuration.

By comparing our accretion rates to that of the results in \citet{Horvath:2005}, we also find that our models of traveling black holes do not substantially increase their mass accretion rate. This leads us to agree with the conclusions of \citet{Custodio:2002} and \citet{Horvath:2005}, that a primordial black hole must sustain a very large boost to begin to accrete sufficient amounts of matter to completely avoid Hawking radiation. When comparing fluid models for $\Gamma= 1.1, 5/4, 4/3$, and $3/2$ in figures \ref{fig:energy_11}, \ref{fig:energy_43}, \ref{fig:energy_54}, and \ref{fig:energy_32}, we find that the stiffer the fluid $\Gamma\rightarrow 2$, the faster the accretion rate. However, the Bondi--Hoyle accretion rates do not appear to increase beyond approximately $5$ to $7$ times the Bondi accretion rate even for ultrarelativistic fluid flows with $v_{\infty}\gtrsim0.9$. We also find that the rotating black hole with spin parameter $a=0$ produces a maximal accretion rate for all models, this is intuitive since the $a=0$ black hole has the largest radius, and consequently the largest ``surface area''. The fluids with adiabatic constant $\Gamma\rightarrow1$ tend to accrete at a rate less than the Bondi accretion rate.

To model more physically realistic accretion in the early universe future models will need to take into consideration the time dependent nature of the background energy density, thus releasing the assumption of a uniform fluid background. Furthermore, it will be beneficial for future studies to consider general relativistic radiation hydrodynamics such as that presented by \citet{Park:2006} which may be used to approximate the radiative loses by the black hole due to Hawking radiation.

\section*{acknowledgments}
The author would like to thank the grant agencies VESF, and the SN2NS project ANR-10-BLAN-0503. We also thank Matt Choptuik for providing both advice and funds from NSERC and CIFAR. We gratefully acknowledge the contributions from the Mathematical Relativity group at the University of Southampton, as well as both the Numerical Relativity and Astrophysics groups at the University of British Columbia. The simulations presented in this article were performed on computational resources supported by WestGrid and Compute/Calcul Canada as well as the PICSciE-OIT High Performance Computing Center and Visualization Laboratory. The author acknowledges the use of the IRIDIS High Performance Computing Facility, and associated support services at the University of Southampton, in the completion of this work. We would also like to thank Ian Hawke for proofreading this paper.
\newpage
\appendix
\section{Code Verification}\label{Appendix:1}
\begin{figure}
\includegraphics[width=3in]{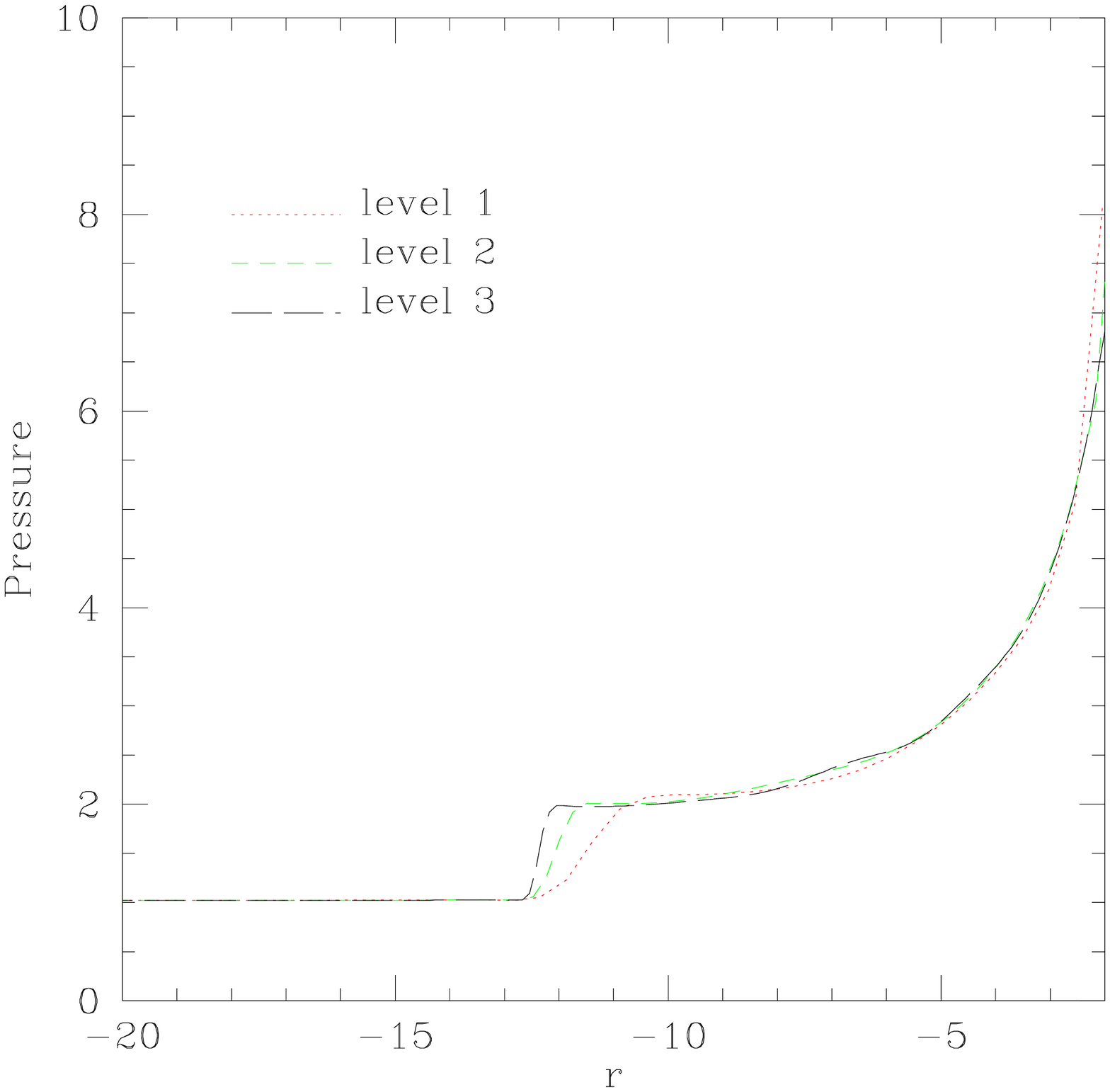}
\caption{A convergence test for model U2. We used a cross-section of the pressure variable in the upstream region (denoted by the negative radial direction) along the axis of symmetry. We see that the system is converging on a shock front upstream of the traveling black hole. Level 1 is $200\times80$, level 2 denotes $400\times160$, and level 3 denotes $800\times320$. In this convergence test we focused on the region $r_{\min}\le r\le 20$ to emphasize the shock front.}\label{fig:v6converge}
\end{figure}

\begin{figure}
\includegraphics[width=3in]{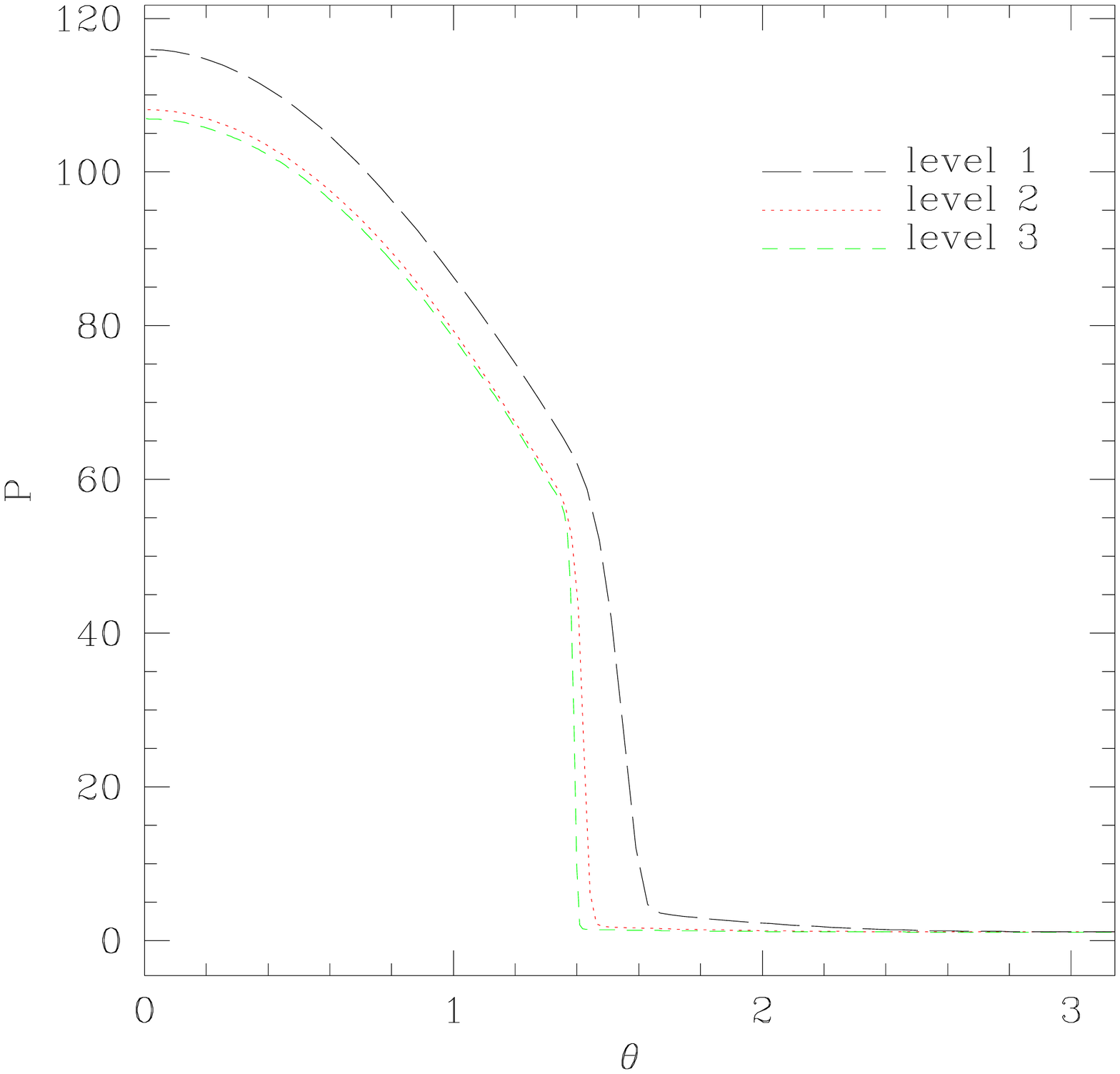}
\caption{A convergence test for model U4. In this test we used a cross section along $r=4$ around the black hole. We see that the solution is converging on a tail shock opening angle of approximately 1.4 radians, and a maximum downstream pressure of approx.~107. Level 1 is $200\times80$, level 2 denotes $400\times160$, and level 3 denotes $800\times320$.}\label{fig:v9converge}
\end{figure}

Since the flow morphology is similar in all cases we present two convergence tests, we show the tests for $v_\infty=0.6$ and $v_\infty=0.9$. These are seen in figures \ref{fig:v6converge} and \ref{fig:v9converge}. The convergence values agree with those used by other authors. The disagreement between the true second order convergence value and my system comes from the special treatment of shock regions and local maxima where the system reduces to first order convergence.

\newpage
\section{Accretion Rates for Different Adiabatic Constants}\label{Appendix:2}
\begin{figure*}
\centering
  \includegraphics[width=3in]{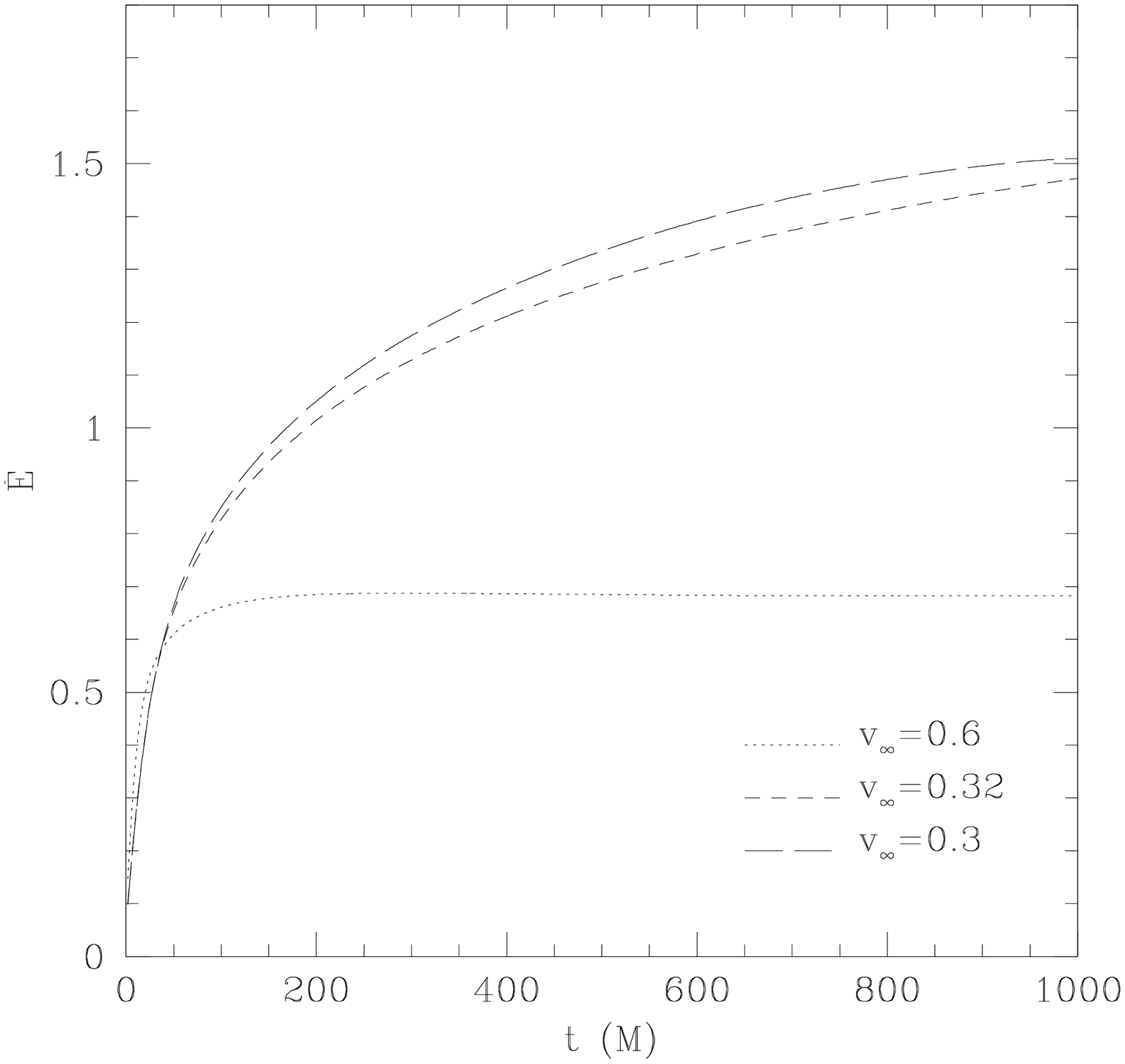}
  \includegraphics[width=3in]{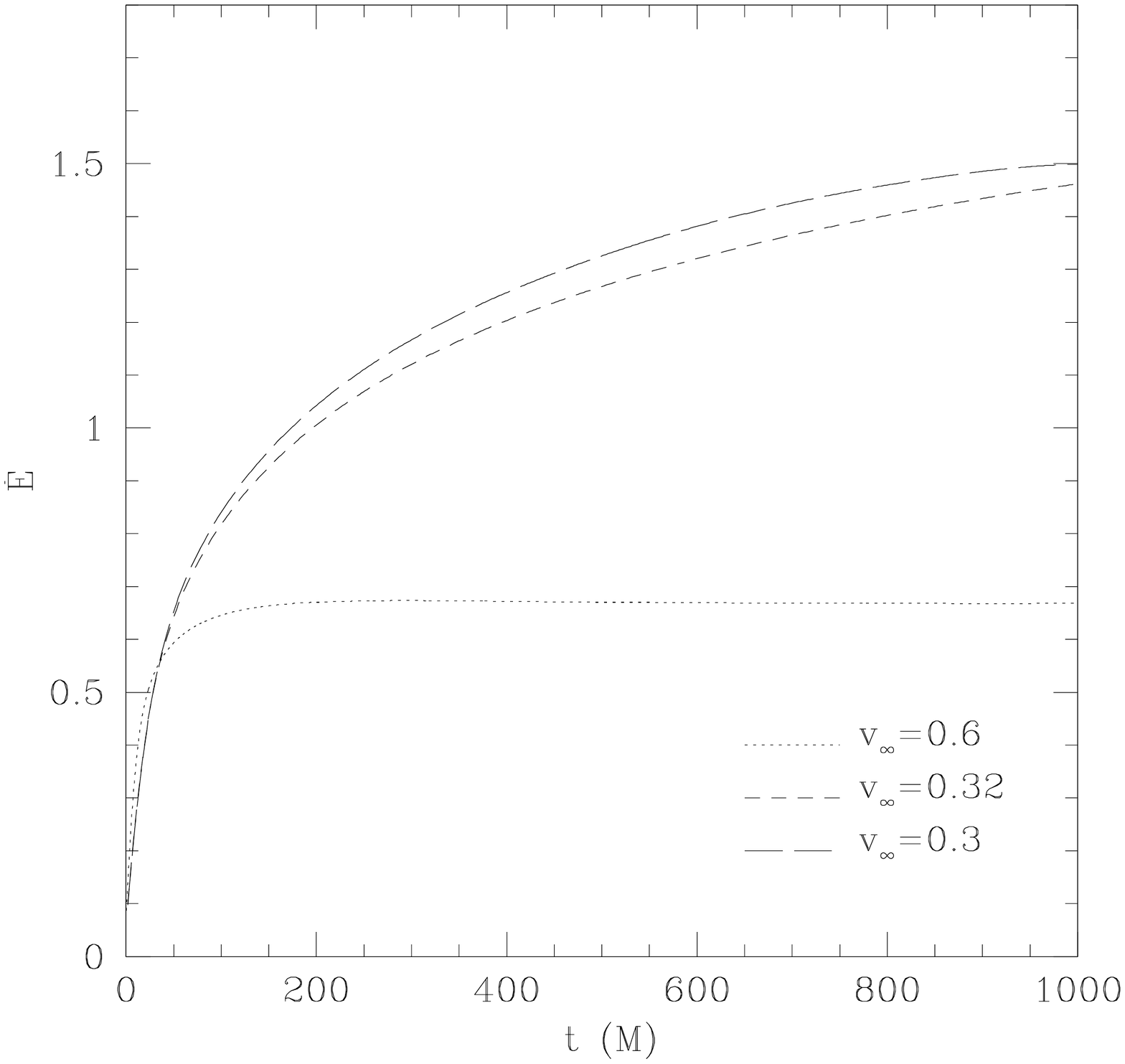}\\
  \includegraphics[width=3in]{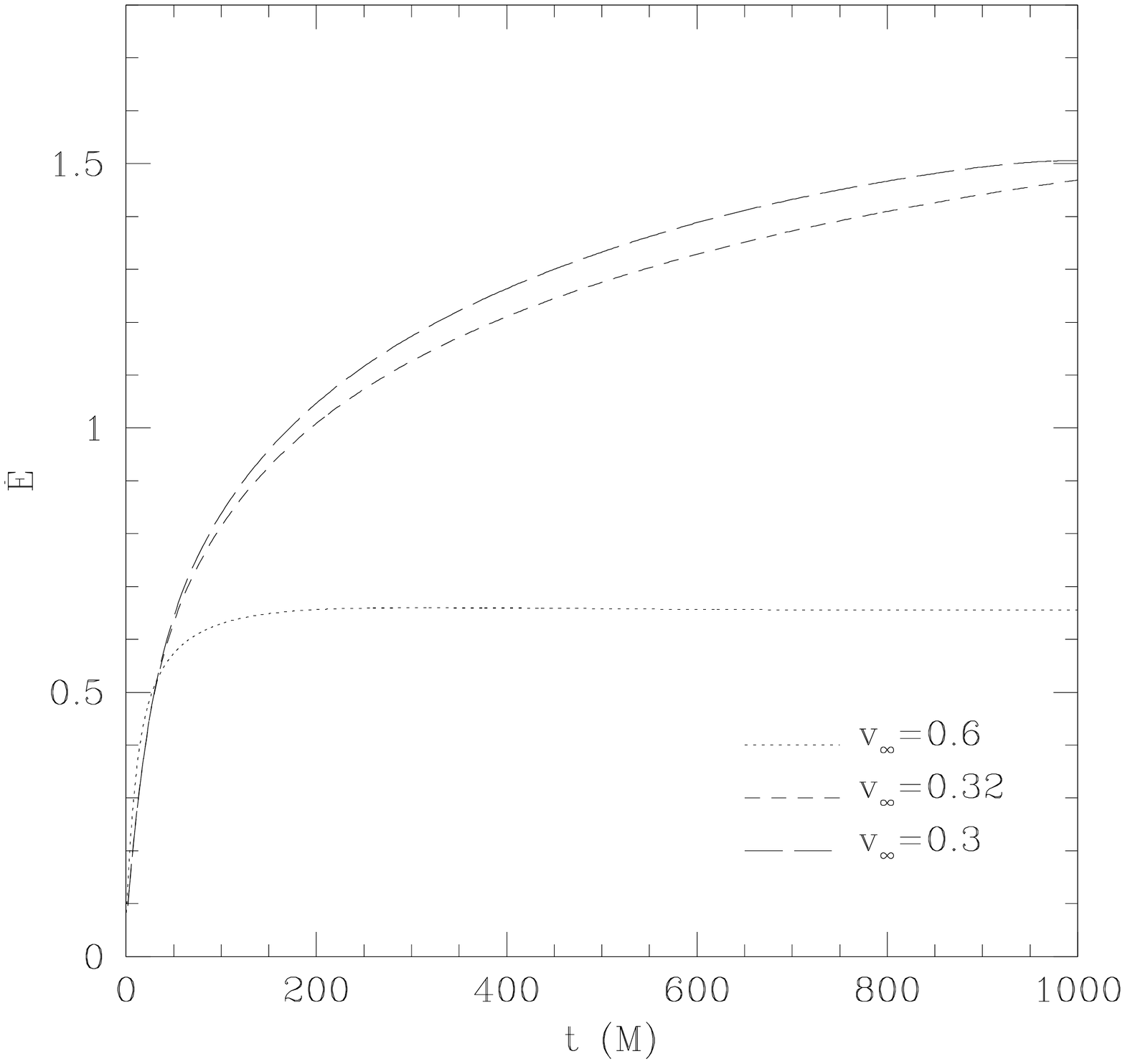}
  \includegraphics[width=3in]{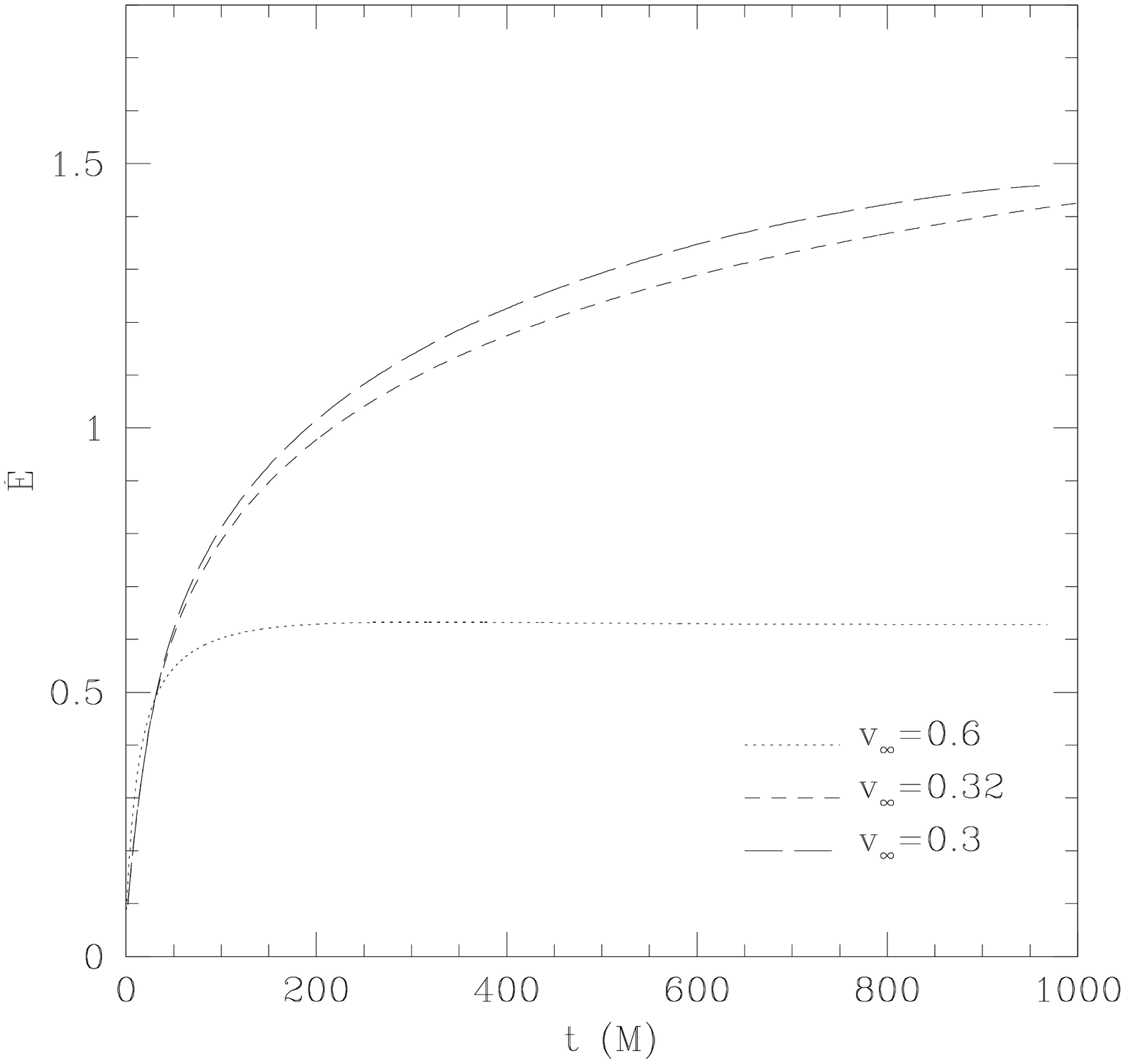}
\caption{The energy accretion rates for models U9 to U11, $\Gamma=1.1$, for all spin parameters. We see that the supersonic flow, $v=0.6$, achieves a steady state solution early in the evolution; however, the subsonic flows have not reached a steady state. To capture the subsonic steady state solution we require the domain of integration to be extended much further than $400M$, and we must run the simulations for longer times.}\label{fig:energy_11}
\end{figure*}

\begin{figure*}
\centering
  \includegraphics[width=3in]{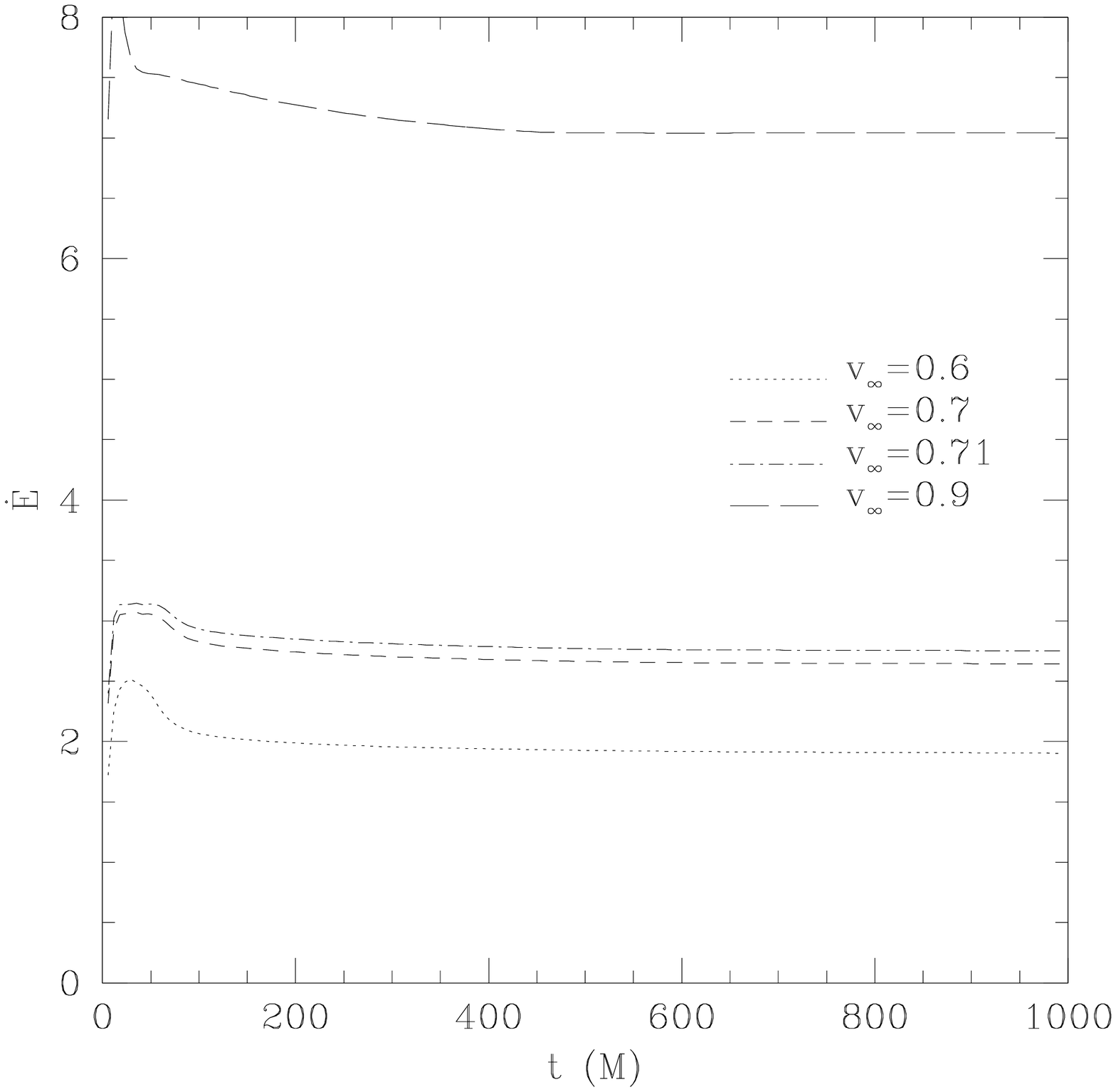}
  \includegraphics[width=3in]{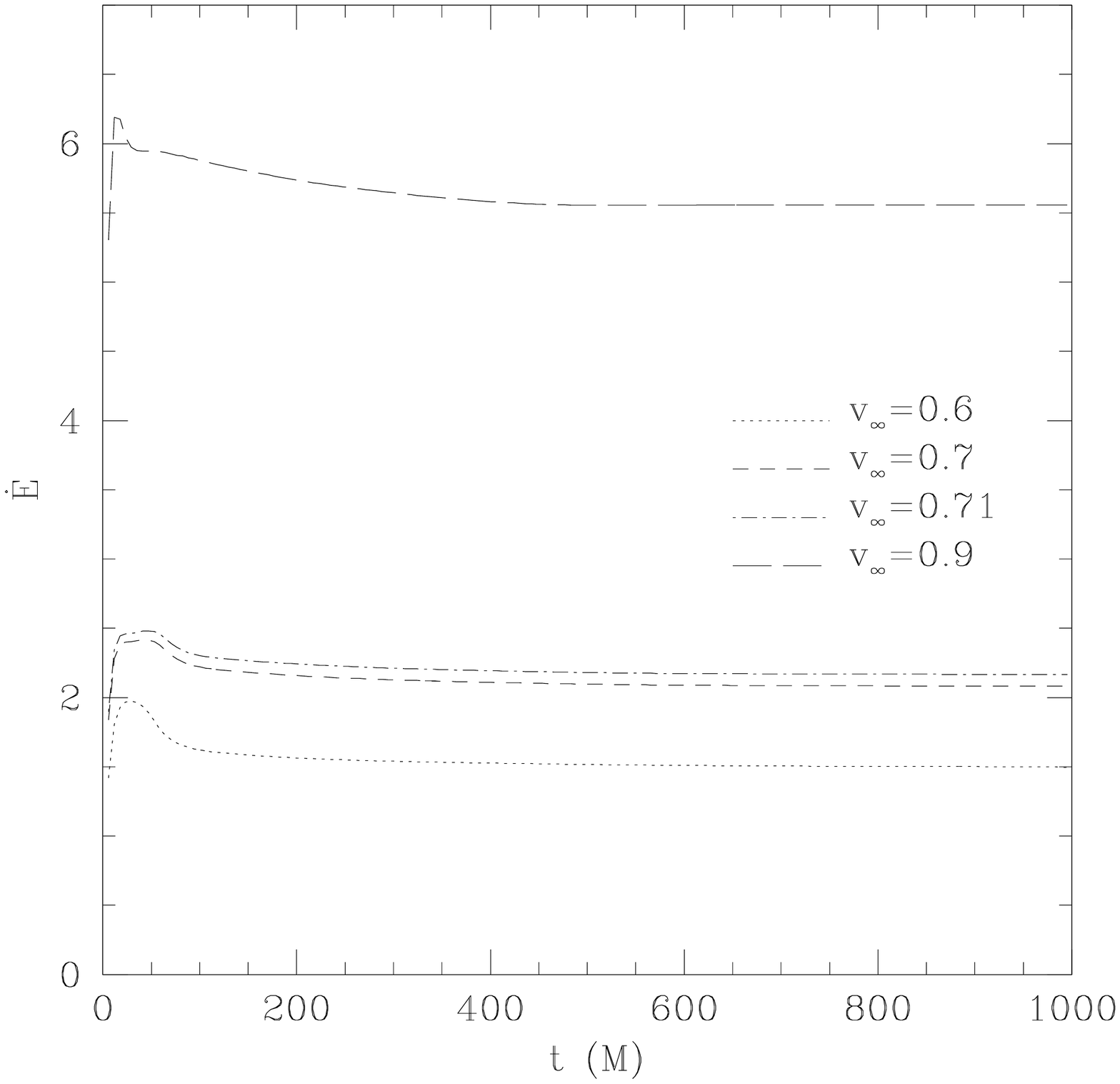}\\
  \includegraphics[width=3in]{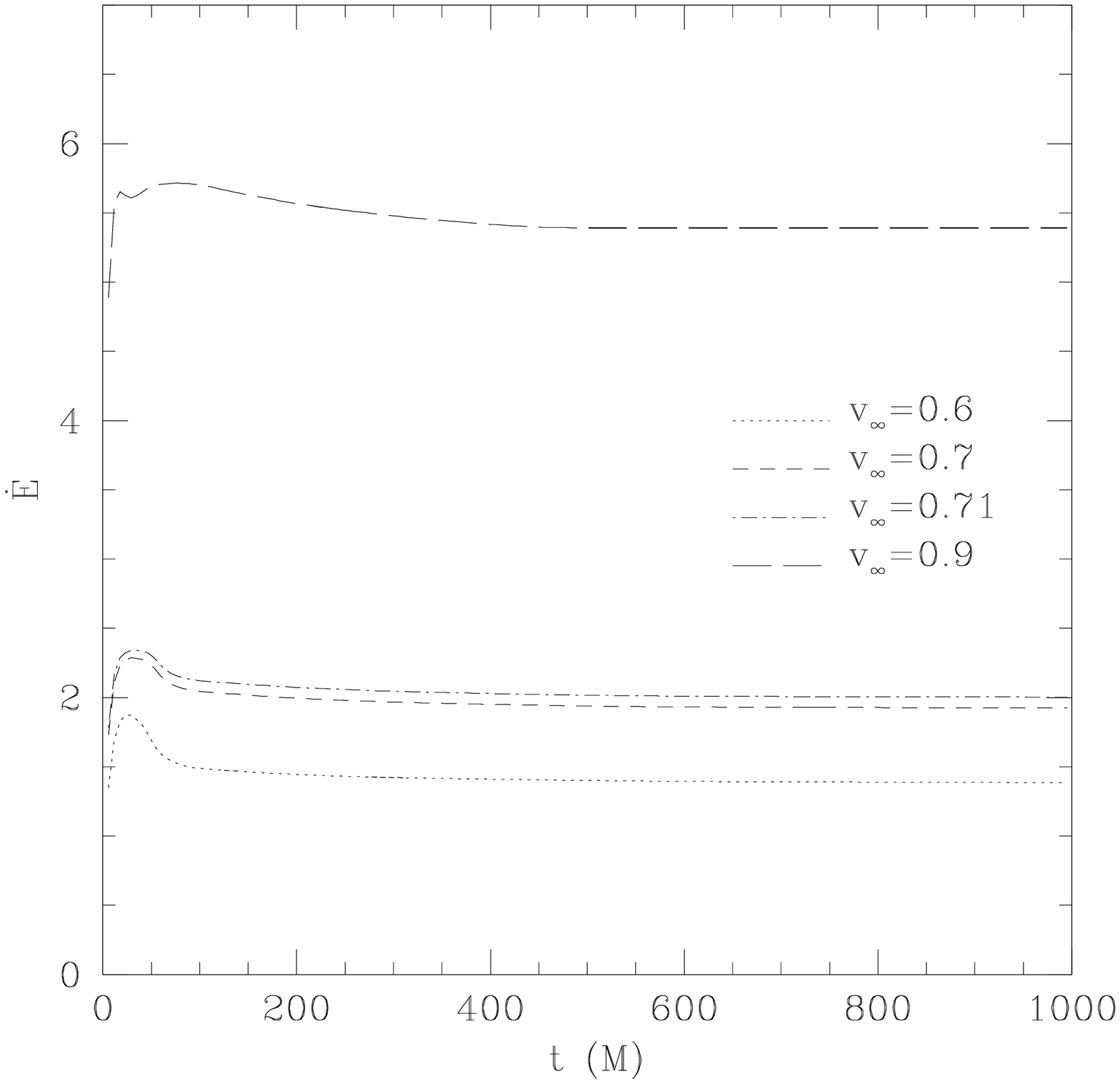}
  \includegraphics[width=3in]{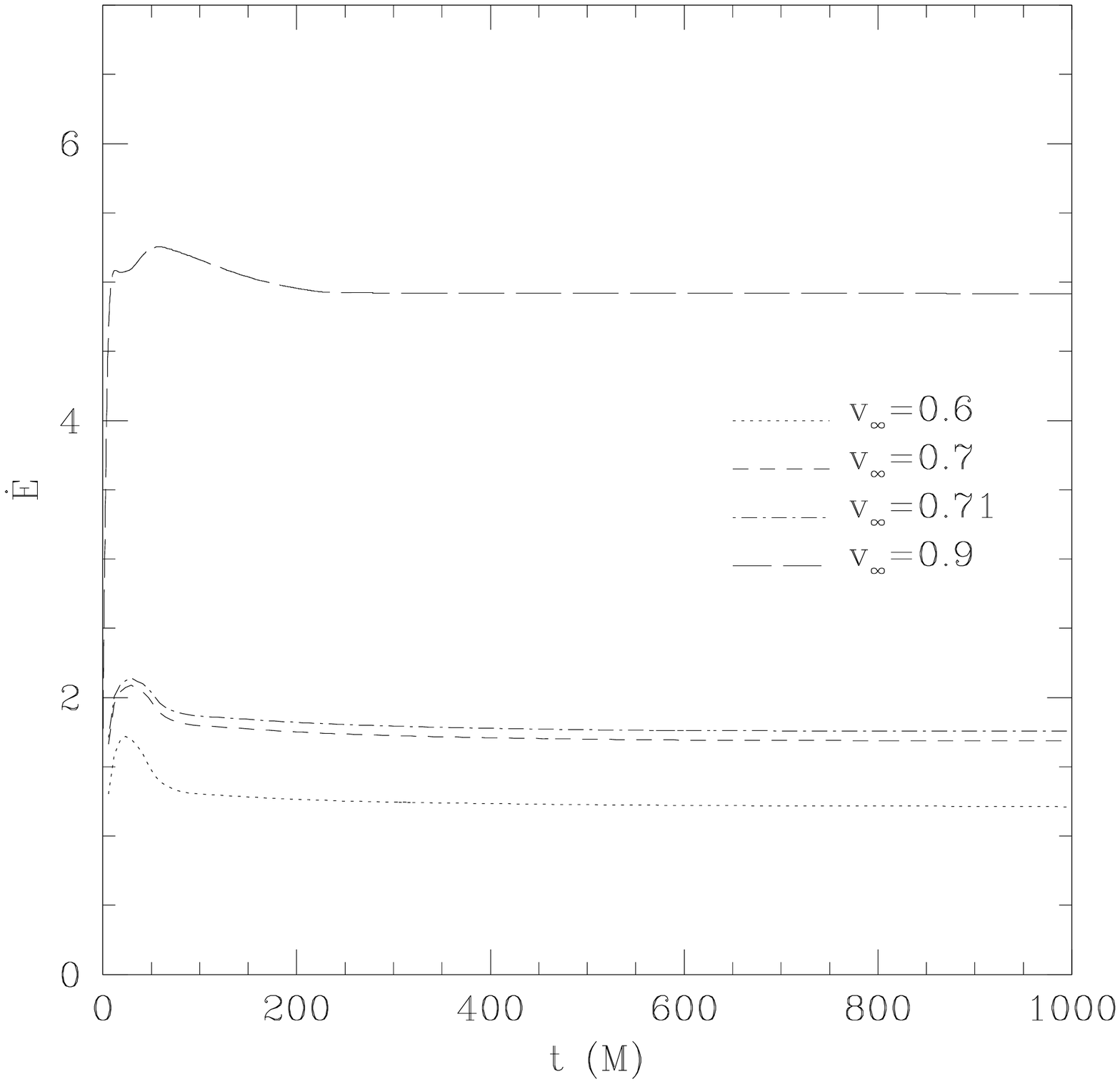}
\caption{The energy accretion rates for models U5 to U8, $\Gamma = 3/2$. This model has a larger accretion rate, and reaches a steady state solution quicker than the softer fluid models.}\label{fig:energy_32}
\end{figure*}

\begin{figure*}
\centering
  \includegraphics[width=3in]{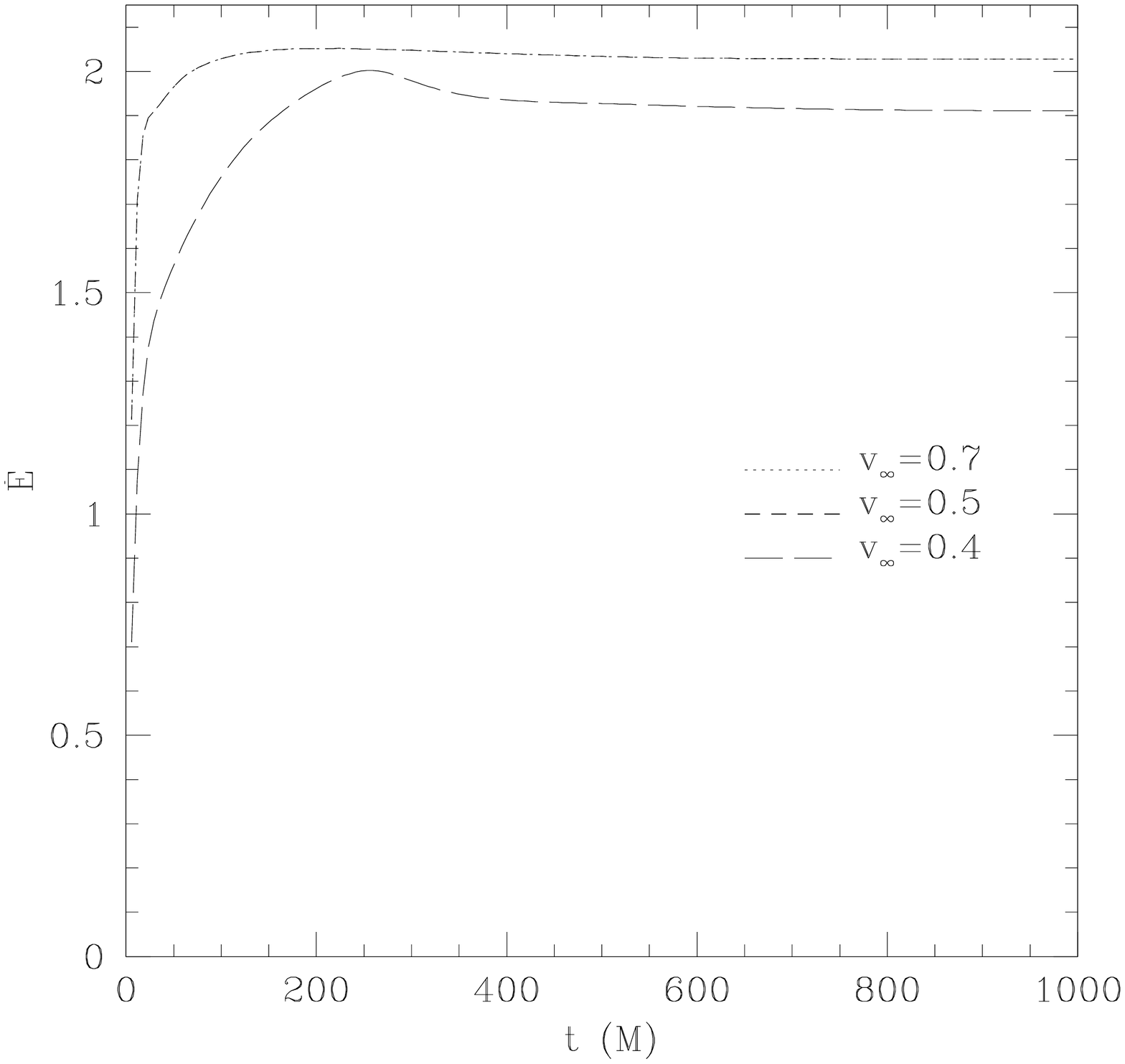}
  \includegraphics[width=3in]{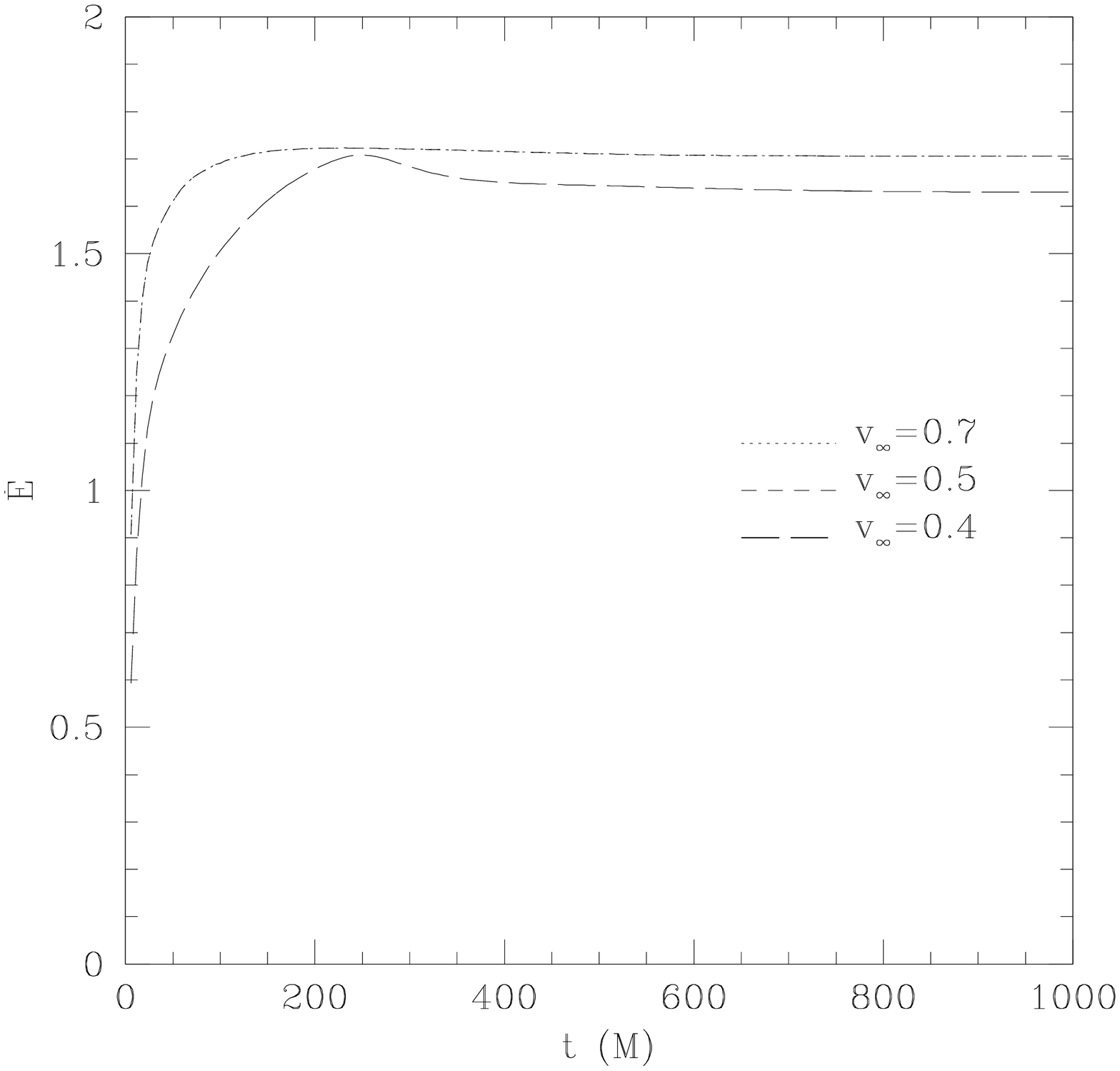}\\
  \includegraphics[width=3in]{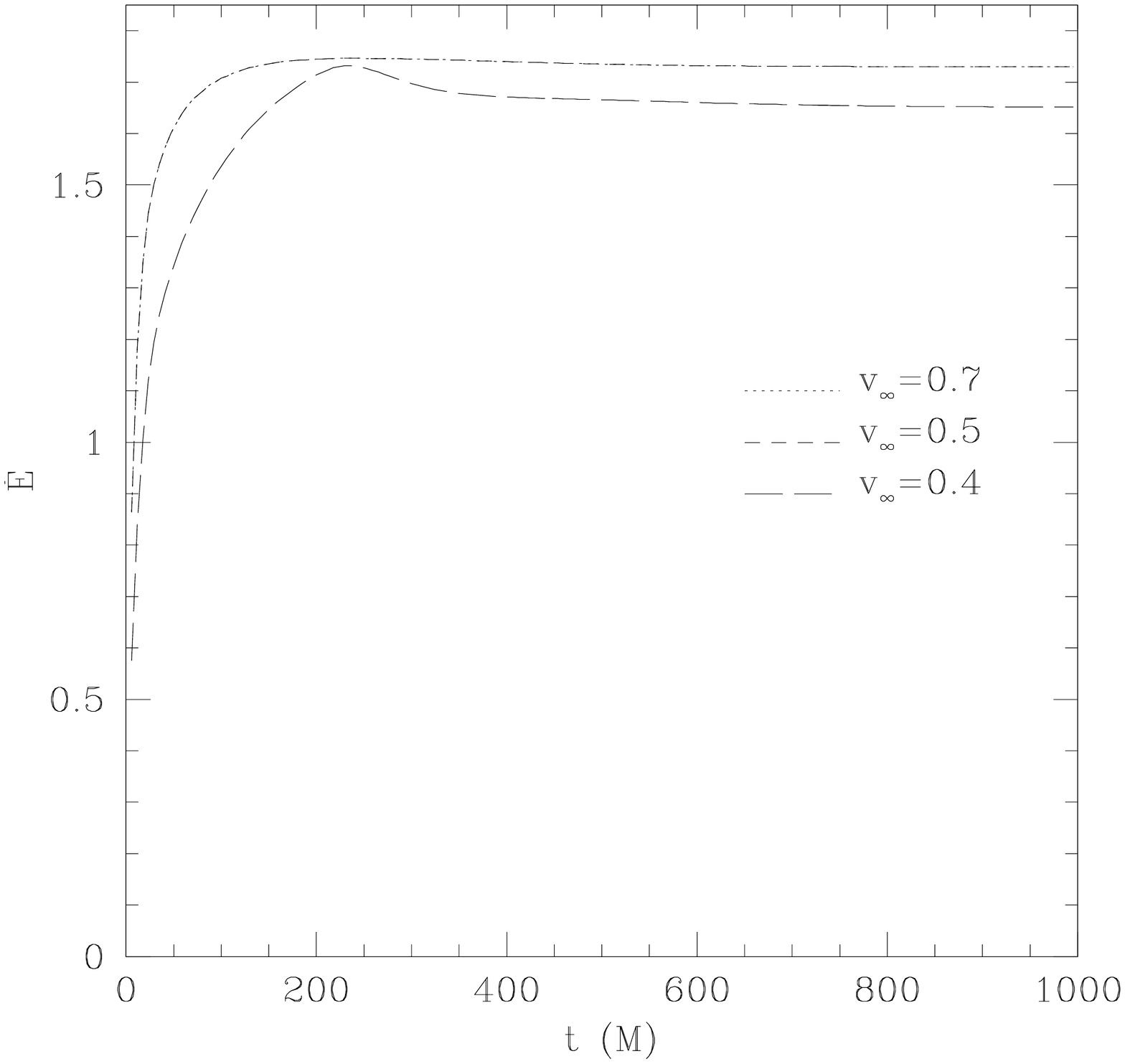}
  \includegraphics[width=3in]{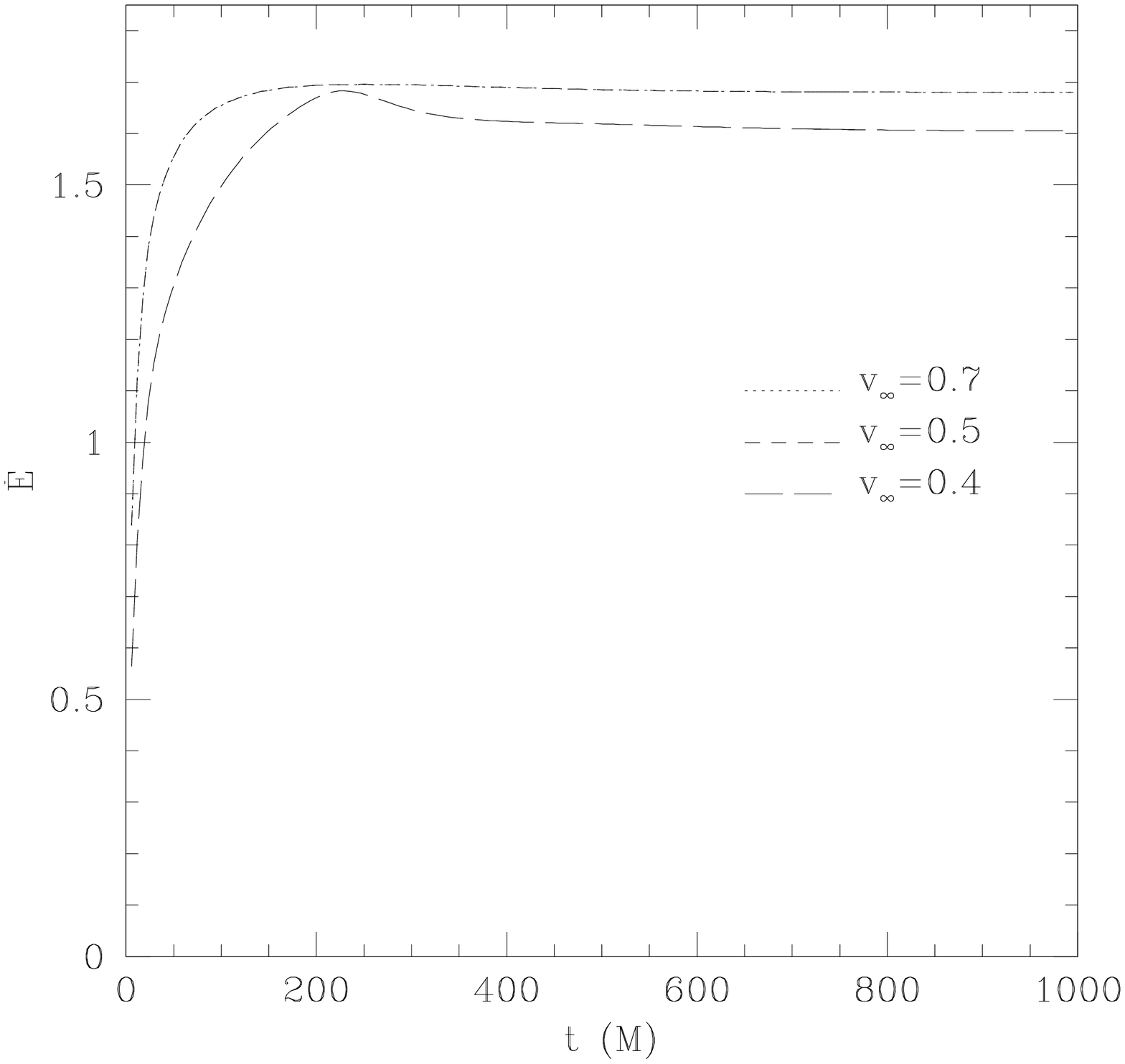}
\caption{The energy accretion rates for models U12 to U14, $\Gamma=5/4$. This model requires a large domain to capture the steady state solution. In agreement with the the models presented in Fig.~\ref{fig:energy_32} this model has a low steady state accretion rate. We note that models U13 and U14 are nearly identical; however, closer inspection reveals that there is a small difference in the accretion rates.}\label{fig:energy_54}
\end{figure*}

In this appendix we present the accretion rates for all the models studied in Table \ref{table:1}. We see that the softer fluid models have a smaller accretion rate and that the subsonic models for $\Gamma=1.1$ have not reached a steady state, as seen in Fig.~\ref{fig:energy_11}. To capture the steady state solution for this model requires a much larger domain of integration, and longer time evolution. As the fluids stiffen the accretion rates increase, while the domain of integration may decrease. This trend is apparent when viewing figures \ref{fig:energy_11}, \ref{fig:energy_32}, \ref{fig:energy_43}, and \ref{fig:energy_54}. 

\bibliographystyle{plainnat}
\bibliography{mybib}

\end{document}